\documentclass[11pt,a4paper]{article}
\usepackage{graphicx}
\usepackage{amsmath}
\usepackage{amsfonts}
\usepackage{amssymb}
\usepackage{amsthm}
\usepackage{color}
\usepackage{tikz}
\usepackage{dsfont}
\usepackage{array}
\usepackage{marginnote}
\usepackage[hidelinks]{hyperref}

\definecolor{mar}{RGB}{65,105,225}
\definecolor{ale}{RGB}{220,20,60}
\definecolor{ian}{RGB}{40,164,40}

\makeatletter
\def\cst#1#2{%
  \ifcsname cst@#1@#2\endcsname%
    #1_{\csname cst@#1@#2\endcsname}%
  \else%
    \ifcsname thecst@#1@@count\endcsname%
    \else%
      \newcounter{cst@#1@@count}%
    \fi%
    \stepcounter{cst@#1@@count}%
    \expandafter\xdef\csname cst@#1@#2\endcsname{\csname thecst@#1@@count\endcsname}#1_{\csname thecst@#1@@count\endcsname}%
  \fi
}
\makeatother

\numberwithin{equation}{section}
\numberwithin{figure}{section}

\newcounter{pointcount}

\makeatletter
\newlength\largearray@width
\def\largearray{\begin{array}{@{}>{\displaystyle}l@{}}\hphantom{\hspace{\largearray@width}}\\[-.5cm]}
\def\endlargearray{\end{array}}
\makeatother

\def\mAthop#1{\displaystyle\mathop{\scriptstyle #1}}

\def\delim#1{%
\noindent%
\leavevmode%
\raise.3em\hbox to\hsize{%
\lower.3em\hbox{\vrule height.3em}%
\hrulefill\ %
\lower.3em\hbox{{\bf #1}}\ %
\hrulefill%
\lower.3em\hbox{\vrule height.3em}}%
\par\penalty10000%
}
\def\enddelim{%
\penalty10000\par\penalty10000%
\noindent%
\leavevmode%
\raise.3em\hbox to\hsize{%
\vrule height.3em%
\hrulefill%
\vrule height.3em}%
\par%
}

\newtheoremstyle{ian}{}{}{\rm}{}{\bf}{}{\newline}{\delim{\thmname{#1} \thmnumber{#2}}\newline\hbox to\textwidth{\hfil\thmnote{\it(#3)}\hfil}}

\theoremstyle{ian}

\newtheorem{theorem}{Theorem}
\newtheorem{lemma}[theorem]{Lemma}
\newtheorem{corollary}[theorem]{Corollary}
\newtheorem{definition}[theorem]{Definition}

\let\oldendtheorem\endtheorem\def\endtheorem{\enddelim\oldendtheorem}
\let\oldendlemma\endlemma\def\endlemma{\enddelim\oldendlemma}
\let\oldendcorollary\endcorollary\def\endcorollary{\enddelim\oldendcorollary}
\let\oldenddefinition\enddefinition\def\enddefinition{\enddelim\oldenddefinition}

\numberwithin{theorem}{section}

\let\a=\alpha

\let\D=\Delta

\let\L=\Lambda

\let\O=\Omega

\let\==\equiv
 
\let\0=\noindent

\def\be{\begin{equation}}
\def\ee{\end{equation}}
\def\bea{\begin{eqnarray}}
\def\eea{\end{eqnarray}}

\begin{document}

\title{\bf Plate-nematic phase in three dimensions}
\author{\bf Margherita Disertori\footnote{Institute for Applied Mathematics \& Hausdorff Center for Mathematics, 
University of Bonn,
Endenicher Allee 60,
D-53115 Bonn, Germany.
E-mail: disertori@iam.uni-bonn.de}, Alessandro Giuliani\footnote{Dipartimento di Matematica e Fisica, Universit\`a degli Studi Roma Tre,
L.go S. L. Murialdo 1, 00146 Roma - Italy, email: giuliani@mat.uniroma3.it}, Ian Jauslin\footnote{Institute for Advanced Study, School of Mathematics, 
1 Einstein Drive,
Princeton, New Jersey,
08540 USA, email: jauslin@ias.edu}}

\vspace{3.truecm}
\date{\today}

\maketitle

\begin{abstract}We consider a system of anisotropic plates in the three-dimensional continuum, interacting via purely hard core interactions. We assume that the particles have a finite 
number of allowed orientations. In a suitable range of densities, we prove the existence of a uni-axial nematic phase, characterized by long range orientational order (the minor axes are 
aligned parallel to each other, while the major axes are not) and no translational order. The proof is based on a coarse graining procedure, which allows us to map the plate model into a 
contour model, and in a rigorous control of the resulting contour theory, via Pirogov-Sinai methods. 
\end{abstract}

\tableofcontents

\section{Introduction}\label{section:introduction}

The mathematical theory of liquid crystalline (LC) phases, even just of their equilibrium properties, 
is still in a primitive stage: most of the predictions on the phase diagram of systems of anisotropic molecules
are based on density functional, or mean field theories. The approximations underlying the derivation of the corresponding effective free energy functionals are typically uncontrolled: 
there is no systematic way of improving the precision, and no rigorous theorem quantifying the error. Ideally, as in any equilibrium statistical mechanics problem, one would like to start 
from a microscopic model of interacting particles, described in terms of (say) a grand-canonical partition function at inverse temperature $\beta$ and activity $z$, and derive bounds on 
the large distance decay of correlations, both for the orientational and the translational degrees of freedom of the particles, for different choices of $(\beta,z)$. Given an inter-particle
interaction, one would like to exhibit values of $(\beta, z)$ at which the correlation functions of the system display broken orientational order and unbroken 
(or partially broken) translational order. Depending on the specific nature of the broken orientational order, and/or of the unbroken/partially broken translational order, 
one names such a phase `uni-axial nematic', or `bi-axial nematic', or `smectic', or `chiral', etc. Essentially none of these phases has ever been mathematically proved to arise in 
any model of interacting particles in the three-dimensional continuum. The purpose of this paper is to report some progress in this direction. Part of our motivation comes from the renewed 
interest of the condensed matter community on the nature of bi-axial nematic phases, which was stimulated by the experimental observation of a bi-axial phase in systems of elongated, 
boomerang-shaped, particles \cite{MDe04, APK04,MKe04}.

\medskip

Let us specify more precisely the context we consider. As is well known, the microscopic interactions responsible for the onset of liquid crystalline phases have electrostatic origin. 
Electrostatic interactions among the microscopic constituents of a liquid crystal are typically strong and repulsive at short distances, and weak and attractive at larger distances (London, 
or Van der Waals, forces). Depending on the specific system under consideration, either the short range repulsion, or the long range attraction, plays a pre-dominant role on the 
onset of the LC phase. It is customary to focus the attention on just one of the two effects, in order to understand which of those is responsible for which LC transitions, if any. 
Of course, if one is after quantitative results, it is important to consider both effects. In this paper, for simplicity, we focus on the effect of repulsive forces, which we model as pure
hard-core interactions. As a consequence, in the model we consider, the temperature plays no role, and the only relevant parameter is the density. 
We also restrict our attention to the case in which the particles have a finite number of allowed orientations, which is a popular, although drastic, simplification. 
It is of great importance to drop this assumption and understand the phenomenon of continuous symmetry breaking in LC, as well as in other, phases of matter. 
We hope to report results in this direction in the future, but this goes beyond the purpose of this paper. 

\medskip

\begin{figure}
  \hfil\includegraphics[width=12cm]{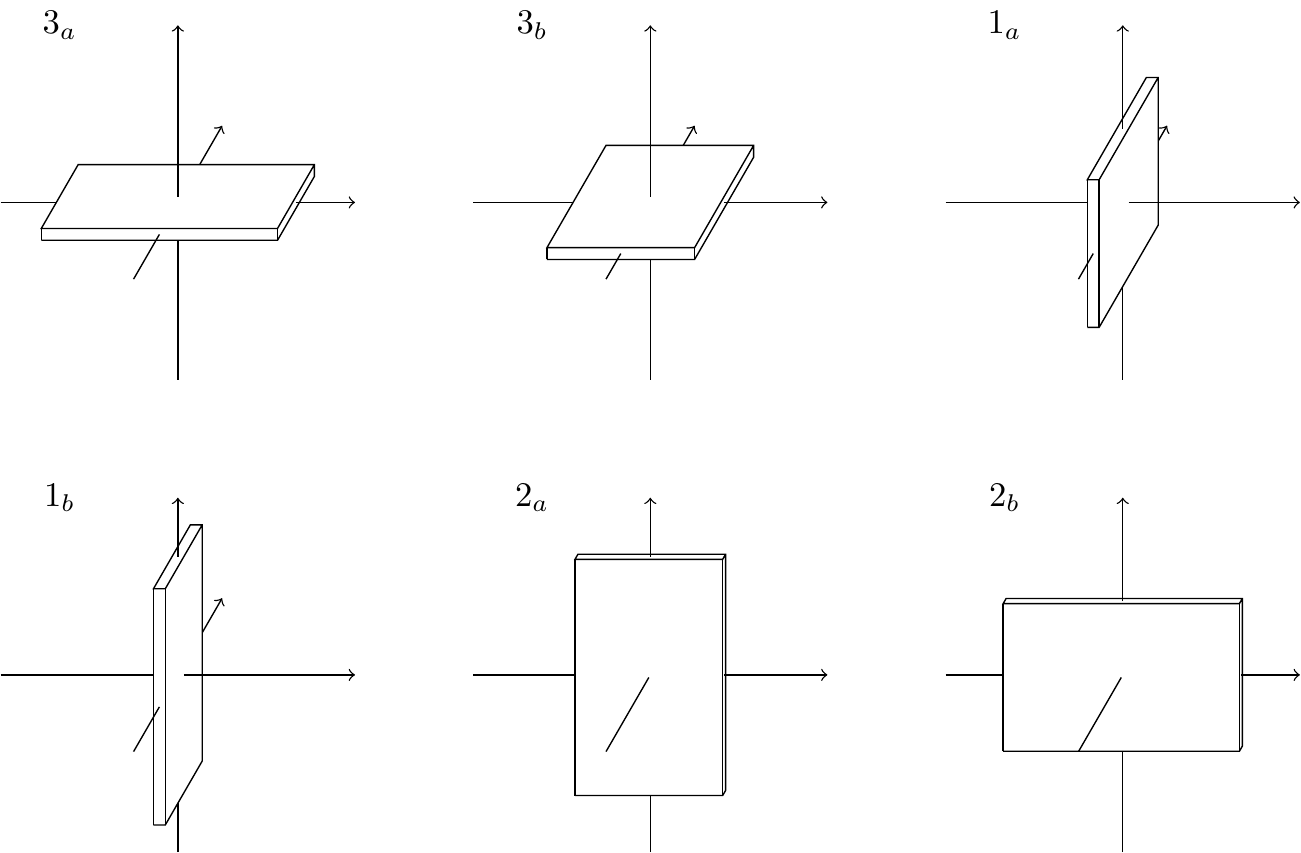}\par\penalty10000
  \caption{The plates and their six allowed orientations}
    \label{figure:plates}
\end{figure}

{\it The model.} Let us now define our model more precisely: we consider a system of hard parallelepipeds of size $1\times k^\alpha\times k$ for some $\alpha\in[0,1]$, which we call 
{\it boards}. If $\alpha<1/2$, a board will be called a {\it rod} and, if $\alpha>1/2$, a {\it plate}. 
The position of each board is given by the position of its center $x\in\mathbb R^3$, and its orientation, which is characterized by a pair of indices $(i,j)\in\{1,2,3\}\times\{a,b\}=:
\mathcal O$ (see Fig.\ref{figure:plates}). 

\begin{figure}
  \hfil\includegraphics[width=12cm]{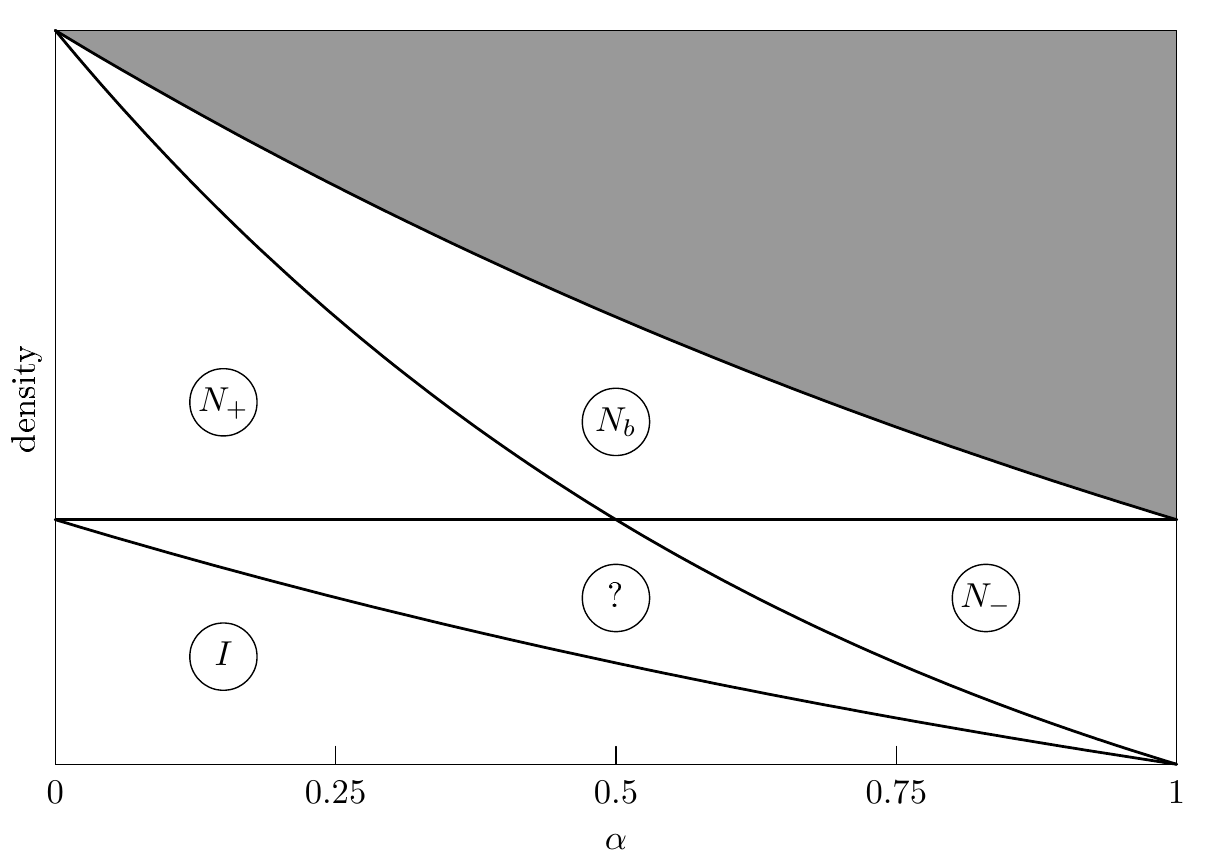}\par\penalty10000
  \caption{Schematic phase diagram for the hard plate model. The phase labeled by $I$ is the {\it isotropic} (no orientational order), $N_-$ is the {\it plate-like nematic} (order in the minor axis), $N_+$ the {\it rod-like nematic} (order in the major axis), and $N_b$ the {\it biaxial nematic} (order in both axes). In the `question mark' region we have no specific prediction about the nature of the phase. 
The region which is grayed out corresponds to densities that are too high for plates to coexist without overlapping.}
  \label{figure:phase}
\end{figure}

We will use the following notation: $(i,j)\equiv i_j$, and a board oriented along $i_j$ will also be said to be ``in the direction $i_j$''. Boards oriented in the direction $i_a$ or $i_b$ will 
be collectively said to be ``of type $i$''. The boards interact via a hard core interaction. We 
shall denote the density of board centers by $\rho$. As the density $\rho$ and the anisotropy exponent $\alpha$ are varied, the system is expected to display a variety of different 
phases, ranging from an isotropic liquid one, to uni-axial and bi-axial nematic, as summarized in Fig.\ref{figure:phase}.

In this paper, we focus on the case of plates, $\alpha>1/2$. For technical reasons, we will restrict to the sub-case $\alpha>3/4$; the significance of the exponent $3/4$ will become 
clear in the course of the proof. Our main result is a rigorous proof of the existence of a uni-axial nematic phase, for $k$ large, $\frac34<\alpha\le 1$, and the density in a suitable,
$(k,\alpha)$-dependent, regime, see below for details. In principle, it should be possible to extend our analysis to smaller values of $\alpha$, most notably to the case of rods. 
It should also be possible to extend it to the case of larger values of the densities, thus substantiating the conjectured existence of a bi-axial nematic phase in our model. 
In both cases, the coarse graining procedure that we employ in the proof is insufficient for a rigorous control of the pressure and correlation functions. 
We hope to report progresses on the phase diagram of the system for more general values of $\rho$ and $\alpha$ in a future publication. 
\par

Before specifying our main results more precisely, let us first give a heuristic idea of why a sequence of transitions from isotropic to nematic phases is expected in our model, as the 
density is increased from zero to its maximum, that is $\rho_{max}=k^{-1-\alpha}$. We focus on the case of very anisotropic plates, $\frac12<\alpha<1$ (the `very' stands for the 
condition that $\alpha<1$). A similar heuristic discussion can be repeated for rods, and is left to the reader. 

Given a plate $(x,o)\in\mathbb R^3\times\mathcal O$, we define the {\it excluded set} on plates of orientation $o'$, as the set of points $y\in\mathbb R^3$ such that 
the plate $(y,o')$ intersects the plate $(x,o)$. The {\it excluded volume} is the volume of the excluded set; it depends on the pair $(o,o')$. If, for instance, $o=3_a$, then 
the excluded volume on plates of different orientations are the following:
\begin{itemize}
  \item the excluded volume produced by $3_a$ on $1_a$ is of the order $k^{2+\alpha}$,
  \item the excluded volume produced by $3_a$ on $1_b$ is of the order $k^{2+\alpha}$,
  \item the excluded volume produced by $3_a$ on $2_a$ is of the order $k^{2+\alpha}$,
  \item the excluded volume produced by $3_a$ on $2_b$ is of the order $k^{1+2\alpha}$,
  \item the excluded volume produced by $3_a$ on $3_b$ is of the order $k^{2}$,
  \item the excluded volume produced by $3_a$ on $3_a$ is of the order $k^{1+\alpha}$,
\end{itemize}
and similarly for the other choices of $(o,o')$. Note that for $k$ large and $\frac12<\a<1$, these excluded volumes are well separated in scales, and ordered as follows: $k^{1+\alpha}\ll k^2\ll k^{1+2\alpha}
\ll k^{2+\alpha}$. This separation of scales, together with the assumption that the distribution of the 
particle centers in space is approximately homogeneous, plays a prominent role in the heuristic explanation of the expected sequence of transitions. 
The sequence of expected transitions can be read from Fig.\ref{figure:phase} above, and is summarized for the reader's convenience here:
\bigskip

\hfil\begin{tikzpicture}

  \draw(-5,0)--(5,0);
  \draw(5,0)++(-0.15,0.15)--++(0.15,-0.15)--++(-0.15,-0.15);
  \draw(5.1,-0.4)node{$\rho$};

  \draw(-3,-0.15)--++(0,0.3);
  \draw(-3,-0.7)node{$\frac1{k^{2+\alpha}}$};
  \draw[line width=0.4mm](-5,-0.2)--++(0,0.4);
  \draw(-5,-0.7)node{$0$};
  \draw(-1,-0.15)--++(0,0.3);
  \draw(-1,-0.7)node{$\frac1{k^{1+2\alpha}}$};
  \draw(1,-0.15)--++(0,0.3);
  \draw(1,-0.7)node{$\frac1{k^{2}}$};
  \draw[line width=0.4mm](3,-0.2)--++(0,0.4);
  \draw(3.5,-0.7)node{$\frac1{k^{1+\alpha}}\equiv \rho_{max}$};

  \draw(-4,0.5)node{$I$};
  \draw(-4,0.5)circle(0.4);
  \draw(-2,0.5)node{$?$};
  \draw(-2,0.5)circle(0.4);
  \draw(0,0.5)node{$N_-$};
  \draw(0,0.5)circle(0.4);
  \draw(2,0.5)node{$N_b$};
  \draw(2,0.5)circle(0.4);
\end{tikzpicture}\par

The letters $I$, $N_-$ and $N_b$ stand for: isotropic phase, uni-axial nematic phase (the $-$ indicates that the minor axes are aligned), and bi-axial nematic phase, respectively. 
The `question-mark' phase has a nature that we cannot establish on the basis of a simple heuristic argument.
The logic behind this conjectured phase diagram is the following. 

\begin{itemize}
  \item Suppose that {$k^{-2}\ll\rho \ll k^{-1-\alpha}$}. Given a plate $(x,o)$, which, without loss of generality, we assume to be in the direction $3_a$ (that is, minor axis along direction $3$ and major axis along direction $1$), there will, typically, be many other plates in the set
\begin{equation}\label{I}
J:=x+\Big\{(y_1,y_2,y_3),\ |y_1|<{\textstyle\frac k2},\ |y_2|<{\textstyle\frac k2},\ |y_3|<\frac{k^\alpha}2\Big\}
\end{equation}
since the volume of $J$ is of the order ${k^{2+\alpha}}$ and $\rho k^{2+\alpha}\gg1$. 

By the hard core constraint, the plates whose centers are in $J$ cannot have orientation $1_a$: their orientations can only be $3_a$, $3_b$, $1_b$, $2_a$ or $2_b$.
In general, plates in different directions can coexist within $J$. However, 
the coexistence is unlikely to happen. In fact, suppose for simplicity that only plates in the directions $3_a$ and $3_b$ coexist within $J$; then, due to the hard core constraint, 
one needs to leave a region of volume $\sim k^{2}$, at the interface between the regions occupied by $3_a$-plates and $3_b$-plates, free of any plate center, an event that 
is very unlikely, since typically any region of such a volume contains many plate centers (because $\rho k^{2}\gg1$). Similarly, if only plates in the directions $3_a$ and $2_b$ coexist, 
one needs to leave a region of volume $\sim k^{1+2\alpha}$, at the interface between the region occupied by $3_a$-plates and $2_b$-plates, free of any plate center, an event that 
is very unlikely, since typically any region of such a volume contains many plate centers (because $\rho k^{1+2\alpha}\gg1$). Analogously, the coexistence between pairs or 
triplets of plates in different directions can be shown to be unlikely. 

Therefore, a typical plate configuration in $J$ consists of many plates, all in the direction $3_a$, with centers distributed approximately uniformly, since their interaction, once we prescribe the 
direction of their axes, is very weak: they `just' have a hard core repulsion that prevents a plate from occupying an excluded region of volume $\sim k^{1+\alpha}$ around the center of 
another plate; on the other hand, any region of volume $k^{1+\alpha}$ within 
$J$ has very small probability of being occupied at all, because $\rho k^{1+\alpha}\ll1$. 

We can now repeat the same argument for a translate $J'$ of $J$ that has intersection of order $k^{2+\alpha}$ with $J$ 
itself: since the intersection typically contains many plates, all oriented in the direction $3_a$, we conclude that also the plates in $J'$ are all in the direction $3_a$, and their centers are 
distributed almost uniformly. Proceeding like this, we conclude, at least heuristically, that the whole space should be covered mostly by plates in the same direction, namely with both the minor and major axes oriented parallel to each other: such a phase is named  {\it biaxial nematic} phase, and denoted by the symbol $N_b$ in the phase diagram. 

  \item Suppose now that {$k^{-1-2\alpha}\ll\rho \ll k^{-2}$}. We proceed as in the previous item: given a plate $(x,o)$, which, without loss of generality, we assume to be in the direction $3_a$, there will, typically, be many other plates in the set  $J$ defined in \eqref{I}. 
By the hard core constraint, as before, the plates whose centers are in $J$ can only be in the directions $3_a$, $3_b$, $1_b$, $2_a$ or $2_b$. However, in order to 
accomodate plates with orientation $1_b$, $2_a$ or $2_b$ within $J$, one needs to leave a region of volume $\sim k^{1+2\alpha}$ free of any plate center, an event that 
is very unlikely, since typically any region of such a volume contains many plate centers, because $\rho k^{1+2\alpha}\gg1$. We conclude that, typically, $J$ contains only 
plates in the directions $3_a$ and $3_b$ (that is, of type $3$): therefore, a typical plate configuration in $J$ consists of many plates, all of type $3$, with centers distributed approximately uniformly, since their interaction, once we prescribe the direction of their minor axes, is very weak: they `just' have a hard core repulsion that prevents a plate from occupying an excluded region of volume $\sim k^{2}$ around the center of another plate; on the other hand, any region of volume $k^{2}$ within 
$J$ has very small probability of being occupied at all, because $\rho k^{2}\ll1$. 

In conclusion, we expect that most of the plates in $J$ are oriented in the directions $3_a$ or $3_b$. Reasoning as in the previous item, we conclude that the whole space 
should be covered mostly by plates of the same type, that is, with their minor axes aligned, a phase named {\it uni-axial}, or {\it plate-like}, nematic ($N_-$) phase.

\item If $\rho k^{2+\alpha}\ll1$, then there are few enough plates that they will almost never be in one another's interaction set, so the system is in an {\it isotropic} ($I$) phase.
\end{itemize}

\medskip

Note that in the list above there is a `gap' in the densities: in fact, in the range $k^{-2-\alpha}\ll\rho \ll k^{-1-2\alpha}$, the reasoning above does not allow us to draw any definite
conclusion about the expected nature of the corresponding phase. As far as we know, even numerically, there is no clear evidence about the existence and nature of long range ordering in this range of densities. 

Another range of densities that is not discussed in the previous list, is the one very close to close-packing. In analogy with the two-dimensional case \cite{GD07}, we expect 
no orientational order at very high densities. It would be very interesting to clarify the (glassy?) nature of the very dense phase, and possibly identify a hidden order parameter 
characterizing its behavior. 

\medskip

We are finally ready to state, informally, our main result. For a more quantitive statement, see the next section. 

\medskip

\noindent{\bf Main result} {\it (informal statement)}: In the context described above, we consider a system of anisotropic plates of size
$1\times k^\alpha\times k$, with $k\gg1$ and $\frac34<\alpha\le 1$, interacting via purely hard-core interactions.
If the density is well within the range where uni-axial nematic ordering is expected, that is, more precisely, if 
$k^{-3\alpha}\log k\ll\rho\ll k^{\alpha-3}$ (note that, for $k$ large, and for the values of $\alpha$ under consideration,
$k^{-1-2\alpha}\ll k^{-3\alpha}\log k$ and $k^{\alpha-3}\le k^{-2}$), then the system is, in fact, in a uni-axial, plate-like, nematic $(N_-)$
phase: in particular, we prove the existence of long range orientational order for the minor axes of the plates, and the absence of
translational order, namely, exponential decay of the truncated center-center correlations. 
\bigskip

\indent The main idea of the proof is to map the model to a coarse-grained contour model and prove that we can compute its partition function and the expectation of local observables using a convergent cluster expansion. To that end, we will first split the lattice $\Lambda$ into cubes of size $\ell:=k/2$. In the range of densities we are studying, each cube contains, on average, many plates (since $\rho k^3\gg1$). We will then define a {\it contour} as the union of cubes that either do not contain one and only one type of plates, or that {\it touch} other cubes which contain plates of a different type. Our endgame will then be to prove that the presence of contours is unlikely, which will imply the main result.

\indent In essence, contours are unlikely because, as we will show below, the probability that a cube contains plates of different types is low. In order to deduce the unlikeliness of contours from the unlikeliness of the  cubes of which it is made, and to control the entropy of the contours, we will use methods coming from the Pirogov-Sinai theory of phase transitions.

The strategy of the proof is very similar to the one of \cite{DG13}, in which a system of hard rods in two dimensions was considered. The main technical novelty lies in the proof that a cube of side $\ell$ containing plates of different types (a `bad cube') has exponentially small 
probability in the big parameter $\rho k^{2+\alpha}$. Once this is proved, the rest of the proof follows closely the one in \cite{DG13} and, therefore, we will not spell out all the details of the proofs, and, instead, refer the reader to~\cite{DG13} in which very similar arguments are expounded. 

As far as we know, our result is the first rigorous one for the onset of a nematic-like phase in systems of finite-size particles, with finite-range interactions, in the three dimensional 
continuum. For previous results, see \cite{AZ86,BKL84,HL79,IVZ06,Ru71,Z96}. We refer to the introduction of \cite{DG13} for a thorough, comparative, discussion of previous results. See also \cite{JL17} for a recent proof of the existence of nematic-like order in a monomer-dimer system with attractive interactions. 

Our inability to rigorously control the bi-axial nematic phase, as well as the optimal range of densities where uni-axial nematic is expected, is related to the highly anisotropic 
shape of the excluded regions created by the hard core interaction around any given plate. For instance, consider the range of densities between $k^{-2}$ and $k^{-1-\alpha}$, where
bi-axial nematic order is expected. From the heuristic discussion above, it would be tempting to think of the `uniformly magnetized' regions, where both the axes of the plates are mostly 
aligned in a common direction, as a union of elementary slabs, each of which is a translate and/or rotation of the region $J$ in \eqref{I}. Even if natural, this choice creates difficulties in 
the treatment of the `transition layers' between different uniformly magnetized regions: these layers, which are the basic constituents of the `Peierls' contours' generically have a wild 
geometric shape, which does not allow us to derive simple bounds on their probability, depending only on their volume. At least, the methods of this paper did not allow us to overcome 
these difficulties: therefore, we limited ourselves to a range of densities where paving the space in cubes allow us to derive effective bounds on the probabilities of the `transition layers', that is, of the connected components of the union of bad cubes.

\section{The model and main results}\label{section:model}

\indent We consider a finite cubic box $\Lambda\subset \mathbb R^3$ of side $L$, such that $L+2{\ell}$ is divisible by $8\ell$, with $\ell:=k/2$. This specific choice is technical and is 
motivated by the definitions of bad regions and contours, see Definitions \ref{def.bad} and \ref{def.con} below; it is conceptually unimportant, since we will eventually send 
$L$ to infinity. We recall that plates are anisotropic parallelepipeds of size $1\times k^\alpha\times k$, with $k\gg 1$ and $\alpha\in(\frac34,1]$, with six possible orientations, as in Fig.
\ref{figure:plates}. We introduce the following notations. Given a plate $p=(x,{o})\in\Lambda\times\mathcal O=:\omega_\Lambda$, let $R_p\subset\mathbb R^3$ denote the geometric 
support of the plate. Given $X\subset\mathbb R^3$, $p$ is said to {\it belong} to $X$ if $x\in X$; $p$ is said to {\it intersect} $X$ ($p\cap X\neq\emptyset$) 
if $R_p\cap X\neq\emptyset$; $p$ is said to be {\it contained} in $X$ if $R_p\subset X$. In addition, given another plate $p'$, $p$ is said to {\it intersect} $p'$ ($p\cap p'\neq\emptyset$) if $R_p\cap R_{p'}\neq\emptyset$.

\indent The {grand canonical partition function of the model} at {\it activity} $z>0$ with open boundary conditions is defined as
\begin{equation}\label{eq:Zopen}
  Z_0(\Lambda)=1+\sum_{n\ge 1}\frac{z^n}{n!} \int_{\omega_\Lambda} dp_1\cdots\int_{\omega_\Lambda} dp_n\ \varphi(p_1,\ldots,p_n)
\end{equation}
where $\int_{\omega_\Lambda} dp$ is a shorthand for $\int_\Lambda dx\sum_{o\in\mathcal O}$, and 
\begin{equation}
  \varphi(p_1,\ldots p_n)=\prod_{i<j}\varphi(p_i,p_j),\qquad \varphi(p,p')=\begin{cases}1\ {\rm if}\  p\cap p'= \emptyset,\\
0\ {\rm if}\  p\cap p'\neq \emptyset.\end{cases}.
\end{equation}
As we shall see below, see the first remark after Theorem~\ref{theorem:nematic}, 
fixing the activity is equivalent to fixing the densities, at least in the range of densities we are interested in. 
\bigskip

In order to prove the main result, we will pick boundary conditions in such a way that one of the types of plates is favored over the others. We will then construct the various infinite volume states by varying the boundary condition. In order to define the boundary condition, we must introduce some additional notation.

We pave $\Lambda$ by cubes of side $\ell$, called ``{\em blocks}'', and by cubes of side $8\ell$, called ``{\em smoothing cubes}'' (since $L+2{\ell}$ is divisible by $8\ell$, the smoothing cubes actually exceed the boundary of $\Lambda$ by 1 block). The lattice of the centers of the blocks is a  lattice of mesh $\ell$, called $\Lambda'$ and the lattice of the centers of the  smoothing cubes is a lattice of mesh $8\ell$, called $\Lambda''$.  Given $\xi\in\Lambda'$, the block centered at $\xi$ is denoted by $\Delta_\xi$, and given $a\in\Lambda''$, the smoothing cube centered at $a$ is denoted by $\mathcal S_a$. Given 
a set $X\subseteq \Lambda$ that is a union of blocks, we denote the {\it coarse-grained} version of $X$ by $X'$:
\begin{equation}
  X=\bigcup_{\xi\in X'}\Delta_\xi.
\end{equation}
We denote the $L_\infty$ distance on $\Lambda$ by
\begin{equation}
  d_\infty((x_1,x_2,x_3),(y_1,y_2,y_3)):=\max\{|x_i-y_i|,\ i\in\{1,2,3\}\}
\end{equation}
and the {\it rescaled} $L_\infty$ distance on $\Lambda'$ by
\begin{equation}
  d'_\infty(\xi,\eta):=\frac{d_\infty(\xi,\eta)}{\ell}.
\end{equation}

We introduce a {\it coarse-grained spin model} on $\Lambda'$: let $\Theta_{\Lambda'}$ denote the set of spin configurations $\sigma\equiv\{\sigma_\xi\}_{\xi\in\Lambda'}$ with $\sigma_\xi\in\{0,1,2,3,4\}$. Given a spin configuration $\sigma\in\Theta_{\Lambda'}$ and a plate configuration $P\in\bigcup_{n\ge0}\omega_\Lambda^n=:\Omega_\Lambda$, $P$ is said to be {\it compatible} with $\sigma$ if it is such that, $\forall\xi\in\Lambda'$,
\begin{itemize}
  \item if $\sigma_\xi=0$, then no plates belong to $\Delta_\xi$,
  \item if $\sigma_\xi=i$ with $i\in\{1,2,3\}$, then every plate that belongs to $\Delta_\xi$ is of type $i$ (this includes the case in which no plates belong to $\Delta_\xi$),
  \item if $\sigma_\xi=4$, then $\Delta_\xi$ contains at least two plates of different type.
\end{itemize}

The set of plate configurations that are compatible with a given spin configuration $\sigma$ is denoted by $\Omega_\Lambda(\sigma)$. In addition, given a block $\Delta_\xi$ and a plate configuration $P$, we denote the restriction of $P$ to $\Delta_\xi$ by $P_\xi$, and we define the set of $P_\xi$'s that are compatible with $\sigma_\xi$ by $\Omega_{\Delta_\xi}^{\sigma_\xi}$ (for example, $\Omega^{1}_{\Delta_\xi}\subset \Omega_{\Delta_\xi}$ is the set of plate configurations in $\Delta_{\xi}$ consisting either of plates of type 1 or of the empty configuration). The subset of $\Omega_{\Delta_\xi}^{\sigma_\xi}$ consisting of configurations with $n$ plates is denoted by $\Omega_{\Delta_\xi}^{n,\sigma_\xi}$.

\bigskip

We rewrite the grand canonical partition function \eqref{eq:Zopen} in $\Lambda$ with open boundary conditions in terms of spin
configurations: 
\begin{equation}
  Z_{0} (\Lambda )= \sum_{ \sigma\in \Theta_{\Lambda'}} \int_{\Omega_\Lambda(\sigma)}
 dP\ \varphi (P)z^{|P|}
\end{equation}
and
\begin{equation}
  \int_{\Omega_\Lambda(\sigma)}dP:=\prod_{\xi\in \Lambda'}\int_{\Omega_{\Delta_\xi}^{\sigma_\xi}}dP_\xi, \quad {\rm with}\quad \int_{\Omega_{\Delta_\xi}^{\sigma_\xi}}dP_\xi=\mathfrak z_0(\sigma_\xi)+\mathds1(\sigma_\xi\neq0)\sum_{n_\xi\ge 1}\frac1{n_\xi!}\int_{\Omega_{\Delta_{\xi}}^{
  n_\xi,\sigma_\xi}}dp_1\cdots d p_{n_\xi}
  \label{dPdef}
\end{equation}
in which $\mathds1(\sigma_\xi\neq0)\in\{0,1\}$ is equal to 1 if and only if $\sigma_\xi\neq0$, and
\begin{equation}
  \mathfrak z_0(1)=\mathfrak z_0(2)=\mathfrak z_0(3)=1,\quad
  \mathfrak z_0(0)=-2,\quad
  \mathfrak z_0(4)=0.\label{eq:2.8}
\end{equation}
The value of $\mathfrak z_0$ is the contribution of the empty configuration to the partition function, and the fact that it equals $-2$ for spin-0 blocks compensates for the fact that the empty configuration is over-counted by $\sigma_\xi=1,2,3$.

We now define the partition function with $q$ boundary conditions, $q\in\{1,2,3\}$, denoted by $Z(\Lambda|q)$:
\begin{equation}
  Z(\Lambda|q)= \sum_{\sigma\in\Theta^q_{\Lambda'}} \int_{\Omega_\Lambda(\sigma)}
  dP\ \varphi (P)z^{|P|}
  \label{Zqdef}
\end{equation}
where $\Theta_{\Lambda'}^q\subset\Theta_{\Lambda'}$ is the set of spin configurations, that are such that $\sigma_\xi=q$ if $d_\infty'(\xi,(\mathbb Z^3\setminus\Lambda)')\le 8$.
The number $8$ appearing here is related to the choice of smoothing cubes of side $8\ell$ and to the fact that $L=8\ell m-2\ell$, for some integer $m$. This specific choice is motivated 
by the definitions of good/bad regions and contours, introduced in Section \ref{section:contour}. In fact, the requirement that 
$\sigma_\xi=q$ if $d_\infty'(\xi,(\mathbb Z^3\setminus\Lambda)')\le 8$ is equivalent to the condition that the `boundary smoothing cubes' (i.e., those intersecting $\L^c$) 
are all good with magnetization $q$, that is, all the sampling cubes that they intersect are good and have magnetization $q$, in the sense of Definition \ref{def.bad}. 
\bigskip

{\bf Remark.} The definition of $Z(\Lambda|q)$ in \eqref{Zqdef} does not require that $\Lambda$ is a cube: it holds in the more general case that $\Lambda$ is a connected region obtained by taking a union of 
smoothing cubes and removing all blocks whose center is at $d_\infty'$-distance equal to 1 from the complement set. 
\bigskip

\indent In the following we will be interested in the {\it $n$-point correlation functions} with $q$-boundary conditions, defined as
\begin{equation}
  \rho_{n}^{(q)}(p_1,\ldots,p_n):=\lim_{\Lambda\nearrow \mathbb R^3} \rho_{n}^{(q,\Lambda)}(p_1,\ldots,p_n)
\label{eq:ncorrf}\end{equation}
with
\begin{equation}
  \rho_{n}^{(q,\Lambda)}(p_1,\ldots,p_n):=\frac1{Z(\Lambda|q)}
  \sum_{\sigma\in\Theta^q_{\Lambda'}}\int_{\Omega_\Lambda(\sigma)}
  dP\ z^{|P|+n}\varphi \big((p_1,\ldots,p_n)\cup P\big)\,.\label{eq:2.11}
\end{equation}

\begin{theorem}[Nematic order]\label{theorem:nematic}
Given $\alpha\in(\frac34,1]$, there exist positive constants $\cst c{smallz},\cst C{bigz},\cst c{expe0},\cst c{expe1},
\cst c{corrdecay}$, such that if $zk^{3-\alpha}\le \cst c{smallz}$ and $zk^{3\alpha}/\log k\ge \cst C{bigz}$, then 
for any $q\in\{1,2,3\}$, $\rho_1^{(q)}(x,o)$ and $\rho_2^{(q)}\big((x_1,o_1),
(x_2,o_2)\big)$ exist and are invariant under translations, that is, $\rho_1^{(q)}(x,o)=\rho_o^{(q)}$ and $\rho_2^{(q)}\big((x_1,o_1),
(x_2,o_2)\big)=\rho_{o_1,o_2}^{(q)}(x_1-x_2)$. Moreover, letting 
\begin{equation}
  \epsilon:=\max\{(zk^2)^{\cst c{expe0}},e^{-\cst c{expe1} zk^{2+\alpha}}\}
  \label{eqepsilon}
\end{equation}
we have, for $m\neq q$,
\begin{equation}
\rho^{(q)}_{q_a}=\rho^{(q)}_{q_b}=z(1+O(\epsilon)),\quad
\rho^{(q)}_{m_a}=\rho^{(q)}_{m_b}=O(z\epsilon) \label{dens}
\end{equation}
and
\begin{equation}
\rho_{o_1,o_2}^{(q)}(x_1-x_2)-\rho^{(q)}_{o_1}\rho^{(q)}_{o_2}=\rho^{(q)}_{o_1}\rho^{(q)}_{o_2}O(\epsilon^{\cst c{corrdecay}|x_1-x_2|/k}).
\label{rho2}
\end{equation}
\end{theorem}
\bigskip

\indent This theorem states that, in the presence of $q$ boundary conditions, most particles are of type $q$ (existence of orientational order), and the truncated two-point correlation function decays exponentially (absence of positional order).
\bigskip

\noindent{\bf Remark}: The proof provides much more detailed information on the set of correlations than what is explicitly stated: in fact, our construction may be applied to the 
computation of the whole set of correlation functions in terms of a convergent cluster expansion, analogous to the one given in Theorem~\ref{theorem:cluster}.
In particular, the equation for the total density as a function of the activity, of the form $\rho=2z(1+O(\epsilon))$, can be inverted via the analytic implicit function theorem, and leads to an equation of the form $z=\frac12\rho(1+O(\epsilon))$: therefore, as anticipated above, fixing the activity or the density is equivalent. We also expect that all higher order density correlations satisfy the cluster property, in analogy 
with the two-point function in \eqref{rho2}, hence the infinite volume Gibbs state with $q$ boundary conditions is pure.
\bigskip

\noindent{\bf Remark}: In order to compute the correlation functions, one can replace the activity $z$ with a plate-dependent activity $\tilde z(p)$ and express the $n$-point truncated correlation function in terms of the partition function with the modified activities:
\begin{equation}
  z^n\left.\frac{\delta^n}{\delta\tilde z(p_1)\cdots\delta\tilde z(p_n)}\log Z(\Lambda|q)\right|_{\tilde z(p)\equiv z}.\label{eq:generating}
\end{equation}
It is, therefore, sufficient to compute the partition function with a plate-dependent activity. 

\subsection{Strategy of the proof}

The proof of our main theorem will be split in several steps, which are summarized here. 
\bigskip

1. We first reformulate the model in terms of contours, which interact via an exponentially decaying potential. The contours arise after coarse graining the hard plate system to the spin model introduced above: the contours can be thought of as the transition layers between different uniformly magnetized regions. The interaction between contours is computed using a cluster expansion.
\bigskip

2. We then map the interacting contour model to a hard core polymer model. In order to compute the pressure of the effective interacting contour model, we perform a Mayer expansion of the multi-contour interaction, to quote D.~Brydges~\cite{Br86}: ``If at first you do not succeed, then expand and expand again''. After this second expansion, the polymer model has a purely hard core interaction.
\bigskip

3. The hard core polymer model can be treated by standard cluster expansion methods, provided the activity of the contours is exponentially small in their size. This is to be expected, because the transition layers contain many {\it bad blocks}, i.e., cubes containing more than one plate orientations. The key technical lemma is a proof that the probability of a single bad block is small, under the assumptions of our main theorem (see Lemma~\ref{lemma:twodircubes}). Building upon this, we obtain the desired estimate on the activity of the contours. A subtle point is that the contour activities are defined inductively, in the spirit of Pirogov-Sinai theory \cite{PS75,KP84} therefore, obtaining the bound on the contour activity from the single-block estimate requires an inductive proof, starting from the smaller contours, and then working our way up to larger ones, which may contain smaller contours in their interior(s). 
\bigskip 

The proof closely follows that in~\cite{DG13}, in which a two-dimensional model of hard rods was considered. The important novelty of the present work is to show that the bad blocks mentioned above are, indeed, unlikely to exist (in~\cite{DG13}, the analogous statement was trivial). For this reason, we will first present, in Section \ref{section:strategy}, the proof that bad blocks are improbable, and then present the remaining arguments, omitting those parts that are mere repetitions of~\cite{DG13}. More precisely, in Section \ref{section:contour} we introduce the contour and hard core polymer representations for the partition function with constant activities, and in Section \ref{section:polymer_cluster} we prove their convergence. Finally, in Section \ref{section:nematic}, we discuss the minor differences arising in the presence of a plate-dependent activity, which, as remarked above, 
is required for the computation of correlation functions, and we explain how to prove the bounds \eqref{dens}-\eqref{rho2}.

\section{Bad blocks and dipoles}
\label{section:strategy}

In this section we prove two basic bounds on the probability of bad blocks (that is, blocks $\Delta_\xi$ with spin $\sigma_\xi=4$)
and bad dipoles (that is, pairs of neighboring blocks $\Delta_\xi,\Delta_\eta$ with spins $\sigma_\xi,\sigma_\eta\in\{1,2,3\}$ and $\sigma_\xi\neq\sigma_\eta$). 

\begin{lemma}[Bad blocks]\label{lemma:twodircubes}
Given a block $\Delta$ which, we recall, is a $k/2\times k/2\times k/2$ cube, let $Z^{\ge 2}(\Delta)$ denote the partition function of plate configurations in $\Delta$ containing at least two different types of plates, and, for $q\in\{1,2,3\}$, let $Z^q(\Delta)$ denote the partition function of type-$q$ plates in $\Delta$:
\begin{equation}
  Z^q(\Delta)=\int_{\Omega_\Delta(q)}dP\ \varphi(P)z^{|P|}.
\end{equation}
There exist positive constants $\cst c{blah},\cst c{expcube},\cst C{blah1}$ such that, if $zk^{3-\alpha}\le \cst c{blah}$ and $zk^{3\alpha}\ge \cst C{blah1}\log k$, then
\begin{equation}
  \frac{Z^{\ge 2}(\Delta)}{Z^q(\Delta)}\le e^{-\cst c{expcube}zk^{2+\alpha}}\;.
  \label{ineqbadcube}
\end{equation}
\end{lemma}
\bigskip

{\bf Proof of Lemma \ref{lemma:twodircubes}.} The main idea of the proof is the following. In the uniformly magnetized system, the block
$\Delta$ contains many plates in the two directions $q_a$ and $q_b$. Whenever two types of plates coexist, there must, because of the hard
core interaction, be a boundary layer between plates of different types, of thickness $k^\alpha$, in which only one of the two directions
is allowed. The volume of this layer is of the order of $k^{2+\alpha}$, which means that the volume that plates can occupy in $Z^{\ge 2}(\Delta)$ is smaller than that in $Z^q(\Delta)$ by $k^{2+\alpha}$. Furthermore, since, as will be shown below, plate partition functions are
exponential in the volume of the available space, this yields a gain factor of order $e^{-zk^{2+\alpha}}$.
\bigskip

\indent In order to estimate the partition functions appearing in this proof, a key tool will be the {\it Mayer expansion}.
The following estimates will often be used. Let $S$ be a subset of $\mathbb R^3$, not necessarily a union of boxes.  
Let  $\Omega^{o}_{S}$, resp.  $\Omega^{q}_{S}$, be the set of plate configurations
of orientation $o\in\{1_a,1_b,2_a,2_b,3_a,3_b\}$, resp. of type $q\in \{1,2,3\}$, and center in $S$; we also denote by 
$\Omega^{n,o}_S$, resp. $\Omega^{n,q}_{S}$, the restriction of $\Omega^{o}_{S}$, resp. $\Omega^{q}_{S}$, to the $n$-plate configurations.   
Then, 
\begin{eqnarray}
&&  \log \int_{\O^o_S}dP\ \varphi(P)z^{|P|}=|S|z(1+O(zk^{1+\alpha})), \label{mayer0.1}\\
&&  \log \int_{\O^q_S}dP\ \varphi(P)z^{|P|}=2|S|z(1+O(zk^2)),
  \label{mayer1}
\end{eqnarray}
where $\int_{\Omega_S^o}dP=1+\sum_{n\ge 1}\frac1{n!}\int_{\Omega_{S}^{n,o}}dp_1\cdots d p_{n}$ and $\int_{\Omega_S^q}dP=1+\sum_{n\ge 1}\frac1{n!}\int_{\Omega_{S}^{n,q}}dp_1\cdots d p_{n}$. This result is a simple extension of the convergence theorems proved for identical particle systems in~\cite{Gr62,Ru63,Pe63}, and follows from the general theory of cluster expansions, discussed at length in many references, among which~\cite{Ru99,Br86,KP86,GBG04}, see also \cite[Section 4.2]{DG13} for a brief introduction.
The factors $zk^{1+\alpha}$ and $zk^{2}$ come from the interaction volumes among plates with the same orientation and same type, respectively. 

\bigskip

The Mayer expansion allows us to estimate the partition function of uniformly magnetized systems, but may not be used whenever several types of plates coexist. To treat this last case, we proceed as follows. 
We split the block $\Delta$ into smaller $k^\alpha/2\times k^\alpha/2\times k^\alpha/2$ cubes, which we call {\it pebbles}. Because of the hard core interaction between plates,
each pebble may only contain plates of a single type. Since $zk^{3\alpha }\gg 1$ each pebble $\delta$ still contains many plates, and the
corresponding partition function can be evaluated by a Mayer expansion: for  $q=1,2,3$ we have by~(\ref{mayer1}),
\[
  Z^q(\delta):=\int_{\O^q_\delta}dP\ \varphi(P)z^{|P|}=e^{\frac14zk^{3\alpha}(1+O(zk^{2}))} 
\]
where we used the fact that the volume of the pebble is $|\delta|=k^{3\alpha }/8$.

Given a configuration of plates in $\Delta$, we color each pebble according to the following.
\begin{itemize}
  \item Every pebble containing at least one plate of type~1 is colored red, of type~2 is colored green, and of type~3 is colored blue.
  \item Empty pebbles are colored black.
\end{itemize}
Every pebble that contains at least {\it two} plates with different orientations is called {\it typical}. Pebbles
that are not typical are called {\it atypical}: every such pebble may be empty, or contain only plates with the same orientation. 
Atypical pebbles owe their name to their low probability: given an atypical pebble $\delta$ and denoting the partition function of atypical configurations in $\delta$ by
$Z^{(\mathrm a)}(\delta):=\sum_{o\in O}\int_{\Omega^o_\delta}dP\ \varphi(P)z^{|P|}-5$, where $5$ compensates the over-counting of empty configurations in the first addend,  
we have, by~(\ref{mayer0.1}),
\begin{equation}
  Z^{(\mathrm a)}(\delta)=6e^{\frac18zk^{3\alpha}(1+O(zk^{2}))}-5. \quad
\end{equation}
Hence
\begin{equation}
  \frac{Z^{(\mathrm a)}(\delta)}{Z^q(\delta)}\le 6e^{-\frac18z k^{3\alpha}(1+O(z k^{2}))}.
  \label{ineqatyp}
\end{equation}
The main idea of the proof is to show that, whenever there are at least two types of plates, $\Delta$ must contain a {\it large} number of atypical pebbles, from which we will prove~(\ref{ineqbadcube}).
\bigskip

Let us show that, if $\Delta$ contains at least two different types of plates, it contains at least $k^{2(1-\alpha)}/2$ atypical pebbles: the proof is based on the two following properties of colorings, which follow from simple geometric considerations:
\begin{itemize}
\item[$(\ast1)$]given a typical pebble of some color, the three $k/2\times k/2\times k^\alpha/2$ tiles that are, respectively, orthogonal to directions~1, 2 
and~3 and contain the pebble, cannot contain a typical pebble of another color.

\item[$(\ast2)$]given a non-empty atypical pebble of some color, at least one of the three $k/2\times k/2\times k^\alpha/2$ tiles that are, respectively, orthogonal to 
directions~1, 2 and~3 and contain the pebble, cannot contain a typical pebble of another color.
\end{itemize}
We will now separately consider the cases in which there is only one color of typical pebbles, and those in which there are at least two (if there are no typical 
pebbles then there are, trivially, $k^{3(1-\alpha)}$ atypical pebbles).

We first consider the case in which there is only one color of typical pebbles. By virtue of the fact that $\Delta$ contains at least two types of plates, there 
exists a non-empty atypical pebble $\delta$ of a different color. By property $(\ast2)$, there is at least one $k/2\times k/2\times k^\alpha/2$ tile containing $\delta$ 
that only contains atypical pebbles, of which there are $k^{2(1-\alpha)}$.

Next, we consider the case in which there are typical pebbles of at least two different colors. We denote the set of typical red, green and blue pebbles by $R_t$, $G_t$ and $B_t$, respectively, and their projection in direction~3 onto 
the lower horizontal face of $\Delta$ (i.e., its `floor') by $r_t$, $g_t$ and $b_t$. By property $(\ast1)$, $r_t$, $g_t$ and $b_t$ are disjoint. We assume, without loss of generality, that $R_t\neq\emptyset$ and $G_t\neq\emptyset$. 
There exists at least one pebble $\delta_r$ in $R_t$ above $r_t$, and by property $(\ast1)$, all the pebbles at the same height as $\delta_r$ that are not above $r_t$ are atypical: therefore, there are at least 
$k^{2(1-\alpha)}-|r_t|$ of them. If $|r_t|\le k^{2(1-\alpha)}/2$, then we 
are done. 
If not, then consider a pebble $\delta_g$ in $G_t$; by property $(\ast1)$, all the pebbles at the same height as $\delta_g$ that are above $r_t$ are atypical: therefore, there are at least $|r_t|>k^{2(1-\alpha)}/2$
of them. 
\bigskip

\indent Now, given a plate configuration $P$, we split
\begin{equation}
  \Delta=\left(\Delta_1^{(\mathrm t)}(P)\cup\Delta_2^{(\mathrm t)}(P)\cup\Delta_3^{(\mathrm t)}(P)\right)\cup\left(\delta_1(P)\cup\cdots\cup\delta_N(P)\right)
\end{equation}
in which $\Delta^{(\mathrm t)}_i(P)$ is the union of typical pebbles of type $i$, and $\delta_j(P)$ is an atypical pebble.
By the discussion above,  $N\ge k^{2(1-\alpha)}/2$, whenever $\Delta$ contains at least two types of plates. We thus have
\begin{equation}
  Z^{\ge2}(\Delta)= \sum_{N=k^{2(1-\alpha)}/2}^{k^{3(1-\alpha)}}\sum_{\displaystyle\mathop{\scriptstyle\underline\Delta^{(t)}\equiv(\Delta_1^{(\mathrm t)},\Delta_2^{(\mathrm t)},\Delta_3^{(\mathrm t)})}_{\underline\delta\equiv(\delta_1,\cdots,\delta_N)}}Z^*(\underline\Delta^{(\mathrm t)},\underline\delta)\;,
  \label{split2}
\end{equation}
in which the sum over $\underline\Delta^{(\mathrm t)}$ and $\underline\delta$ is the sum over subsets for which there exists a plate configuration $P$ such that $\underline\Delta^{(\mathrm t)}\equiv\underline\Delta^{(\mathrm t)}(P)$ and $\underline\delta\equiv\underline\delta(P)$, and $Z^*(\underline\Delta^{(\mathrm t)},\underline\delta)$ is the partition function of plate configurations $P$ such that $\underline\Delta^{(\mathrm t)}(P)\equiv\Delta^{(\mathrm t)}$ and $\underline\delta(P)\equiv\underline\delta$. Furthermore,
\begin{equation}\label{ab0}
  Z^*(\underline\Delta^{(\mathrm t)},\underline\delta)\le
  \left(\prod_{i=1}^{3}Z^i(\Delta^{(\mathrm t)}_i)\right)
  \left(\prod_{j=1}^{N}Z^{(\mathrm a)}(\delta_j)\right)
\end{equation}
and, by~(\ref{mayer1}) with $|S|=|\Delta^{(\mathrm t)}_i|\leq k^{3}/8$, it holds for every $i,q\in\{1,2,3\}$,
\begin{equation}\label{ab1}
  Z^i(\Delta_i^{(\mathrm t)})=
  Z^q(\Delta_i^{(\mathrm t)})e^{O(zk^3zk^2)}.
\end{equation}
We now split the denominator $Z^q(\Delta)$, which, we recall, is the partition function with at most one type of plates. By~(\ref{mayer1}),
\begin{equation}\label{ab2}
  Z^q(\Delta)=
  \left(\prod_{i=1}^{3}\left(Z^q(\Delta_i^{(\mathrm t)})e^{O(zk^3zk^2)}\right)\right)
  \left(\prod_{j=1}^{N}\left(Z^q(\delta_j)e^{O(zk^{3\alpha}zk^2)}\right)\right).
\end{equation}
Thus, by~(\ref{ineqatyp}), \eqref{ab0}, \eqref{ab1}, \eqref{ab2}, 
\begin{equation}
  \frac{Z^*(\Delta^{(t)},\underline\delta)}{Z^q(\Delta)}
  \le
  e^{O(zk^3zk^2)}\,
  6^Ne^{-\frac{N}8zk^{3\alpha}(1+O(zk^2))}
\end{equation}
which we plug into~(\ref{split2}), thus getting
\begin{equation}\label{eq.3.13}
    \frac{Z^{\ge2}(\Delta)}{Z^q(\Delta)}
    \le
    e^{O(zk^3zk^2)}
    \sum_{N=k^{2(1-\alpha)}/2}^{k^{3(1-\alpha)}}
    {k^{3(1-\alpha)}\choose N}\,
    e^{-\frac{N}8zk^{3\alpha}(1+O(zk^2))}(6C)^N,\end{equation}
for some constant $C>0$. Here $C^N{k^{3(1-\alpha)}\choose N}$ is an upper bound for the number of terms in the sum over $\underline\Delta^{(t)}$ and $\underline\delta$. To see this, an observation that plays a key role is that 
$\Delta_1^{(\mathrm t)},\Delta_2^{(\mathrm t)}$ and $\Delta_3^{(\mathrm t)}$ must be mutually disconnected, because of property $(*1)$ [note: two pebbles that touch at an edge or a corner are considered as disconnected].  Therefore, each connected component of 
$\Delta^{(t)}=\Delta_1^{(\mathrm t)}\cup \Delta_2^{(\mathrm t)}\cup\Delta_3^{(\mathrm t)}$ is either of type 1, or 2, or 3; moreover, it must be adjacent to at least one atypical pebble, 
which implies that the number of connected components of $\Delta^{(t)}$ is certainly smaller than $6N$ (here $6$ is the number of faces of an atypical pebble).
Given these observations, it is easy to count the number of 
terms in the sum over $\underline\Delta^{(t)}$ and $\underline\delta$: in fact, we can first choose the atypical pebbles, which costs a factor smaller or equal than ${k^{3(1-\alpha)}\choose N}$, and then sum over the partitions of $\Delta^{(t)}$ into the three sets $\Delta_1^{(\mathrm t)},\Delta_2^{(\mathrm t)},\Delta_3^{(\mathrm t)}$. Such a sum over partitions costs at 
most a factor $3^{N'}$, where $N'$ is the number of connected components of $\Delta^{(t)}$, and $3$ is the number of `colors' (that is, $1,2$ or $3$) that we can attach to 
each connected component. As observed above, $N'\le 6N$, so that the constant $C$ in \eqref{eq.3.13} is smaller than $3^6$. 
From \eqref{eq.3.13} we immediately get:
 \begin{eqnarray}   
    \frac{Z^{\ge2}(\Delta)}{Z^q(\Delta)}
  &  \le&     e^{O(zk^3zk^2)}  e^{-\frac1{32} k^{2(1-\alpha)}\cdot zk^{3\alpha}(1+O(zk^2))}
    \sum_{N=0}^{k^{3(1-\alpha)}}
    {k^{3(1-\alpha)}\choose N}\,
    e^{-\frac1{16}Nzk^{3\alpha}(1+O(zk^2))}(6C)^N\nonumber\\
&=&     e^{O(zk^3zk^2)}  e^{-\frac1{32} zk^{2+\alpha}(1+O(zk^2))}\Big(1+6C e^{-\frac1{16}zk^{3\alpha}(1+O(zk^2))}\Big)^{k^{3(1-\alpha)}}\\
&\le&\exp\left(
      -\frac1{32}zk^{2+\alpha}
      \left(
	1+O(zk^{3-\alpha})+O(z^{-1}k^{1-4\alpha}e^{-\frac1{17}zk^{3\alpha}})\right),
    \right)
\end{eqnarray}
where the exponent $\frac1{17}$ in the last line may be replaced by any exponent smaller than $\frac1{16}$, for $zk^2$ sufficiently small. 
The last term can be bounded as follows $z^{-1}k^{1-4\alpha}e^{-\frac1{17}zk^{3\alpha}}=
\frac{1}{zk^{3\alpha }}k^{1-\alpha } e^{-\frac1{17}zk^{3\alpha}}\ll k^{1-\alpha } e^{-\frac1{17}zk^{3\alpha}}.$
This, provided $zk^{3\alpha}\gg\log k$ and $zk^{3-\alpha}\ll1$, implies~(\ref{ineqbadcube}).\qed
\bigskip

\begin{corollary}[Bad dipoles]\label{corollary:twodircubes}
Given two blocks $\Delta_1$ and $\Delta_2$ that have a common face, let $Z^{\ge2}(\Delta_1\cup\Delta_2)$ denote the partition function of
plates in $\Delta_1$ and $\Delta_2$, that are such that $\Delta_1$ and $\Delta_2$ are uniformly magnetized and have different magnetizations. 
There exist positive constants $\cst c{blah.0},\cst c{expdipole},\cst C{blah1.0}$ such that, if $zk^{3-\alpha}\le \cst c{blah.0}$ and $zk^{3\alpha}\ge \cst C{blah1.0}\log k$, then
\begin{equation}
  \frac{Z^{\ge2}(\Delta_1\cup\Delta_2)}{Z^q(\Delta_1\cup\Delta_2)}\le e^{-\cst c{expdipole}zk^{2+\alpha}}.
  \label{ineqbaddipole}
\end{equation}
\end{corollary}

{\bf Proof of corollary~\ref{corollary:twodircubes}.} Consider the $k/2\times k/2\times k/2$ cube $\Delta$ that has half its volume in $\Delta_1$ and half in $\Delta_2$. 
Without loss of generality, we assume that $\Delta_1$ is to the left of $\Delta_2$ in direction 1. Since $\Delta_1$ and $\Delta_2$ are uniformly magnetized and have different magnetizations, the plate configuration restricted to 
the central cube $\Delta$ either has two plates of different types, or is at least half empty. In the second case, we may assume without loss of generality that all plate centers belong to the left half of the cube. Therefore, 
we bound
\begin{eqnarray}
  Z^{\ge 2}(\Delta_1\cup\Delta_2)&\le&
  \sum_{\substack{1\le i,j\le 3\, :\\ i\neq j}}Z^i(\Delta_1\setminus\Delta)\Big[2Z^{1,\emptyset}(\Delta)+Z^{\ge 2}(\Delta)\Big]Z^j(\Delta_2\setminus\Delta)\\
  &=&6\,Z^q(\Delta_1\setminus\Delta)\Big[2Z^{1,\emptyset}(\Delta)+Z^{\ge 2}(\Delta)\Big]Z^q(\Delta_2\setminus\Delta)e^{O(zk^3zk^2)},
\end{eqnarray}
where $Z^{1,\emptyset}(\Delta)$ is the partition function of plate configurations in $\Delta$, such that the plates are all of the same type, and the right half of the box is empty, i.e., 
contains no plate centers. 
Using (\ref{mayer1}), we find 
\begin{equation}
Z^{1,\emptyset}(\Delta)=3Z^q(\Delta)e^{-\frac1{8}zk^3+O(zk^3zk^2)}\;
\end{equation}
and 
\begin{equation}
  Z^q(\Delta_1\cup\Delta_2)=Z^q(\Delta_1\setminus\Delta)Z^q(\Delta)Z^q(\Delta_2\setminus\Delta)e^{O(zk^3zk^2)}.
\end{equation}
The corollary then follows directly from these two equations and from Lemma~\ref{lemma:twodircubes}.\qed

\section{The contour theory.}\label{section:contour}

\indent In this section, we construct first the interacting contour model and then the hard core polymer system that were mentioned above. The basic idea of the construction of the 
contours is that a spin configuration can be seen as a union of connected `uniformly magnetized regions' (union of blocks that all have the same spin, equal to $q\in\{1,2,3\}$), and 
boundary regions where either the spin changes, or there are `defects' (blocks with spin equal to $0$ or $4$). 
Contours will be defined as structures that comprise information about the location and nature of these boundaries, as well as the value of the spin on either side 
of the boundary. To make this precise, we must first {\it locate} the boundary, which we do by defining the concepts of `good' and `bad' regions.


\subsection{Goodness, badness and contours}\label{subsection:goodness}

\begin{figure}
  \hfil\includegraphics[width=3cm]{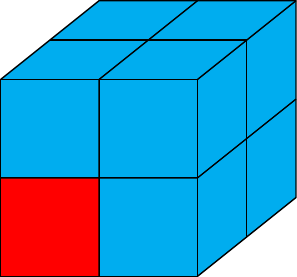}
  \caption{The sampling cube associated to the red (color online) block.}
  \label{figure:sampling}
\end{figure}

\begin{definition}[Sampling cubes]
Given $\xi\in\Lambda'$, we define the {\it sampling cube} associated to $\xi$ as
\begin{equation}
  S_\xi= \bigcup_{\mAthop{\eta\in\Lambda'}_{0\le\eta_i-\xi_i\le\ell}}\Delta_\eta
\end{equation}
where $\xi_i$ and $\eta_i$, $i=1,2,3$, are the coordinates of $\xi,\eta\in\Lambda'$ (see Fig.~\ref{figure:sampling}). Note that if $d'_\infty(\xi,\Lambda_c')>1$, then $S_\xi$ contains exactly {\em 8 blocks}.
\end{definition}
\bigskip

\begin{figure}
  \hfil\includegraphics[height=9cm]{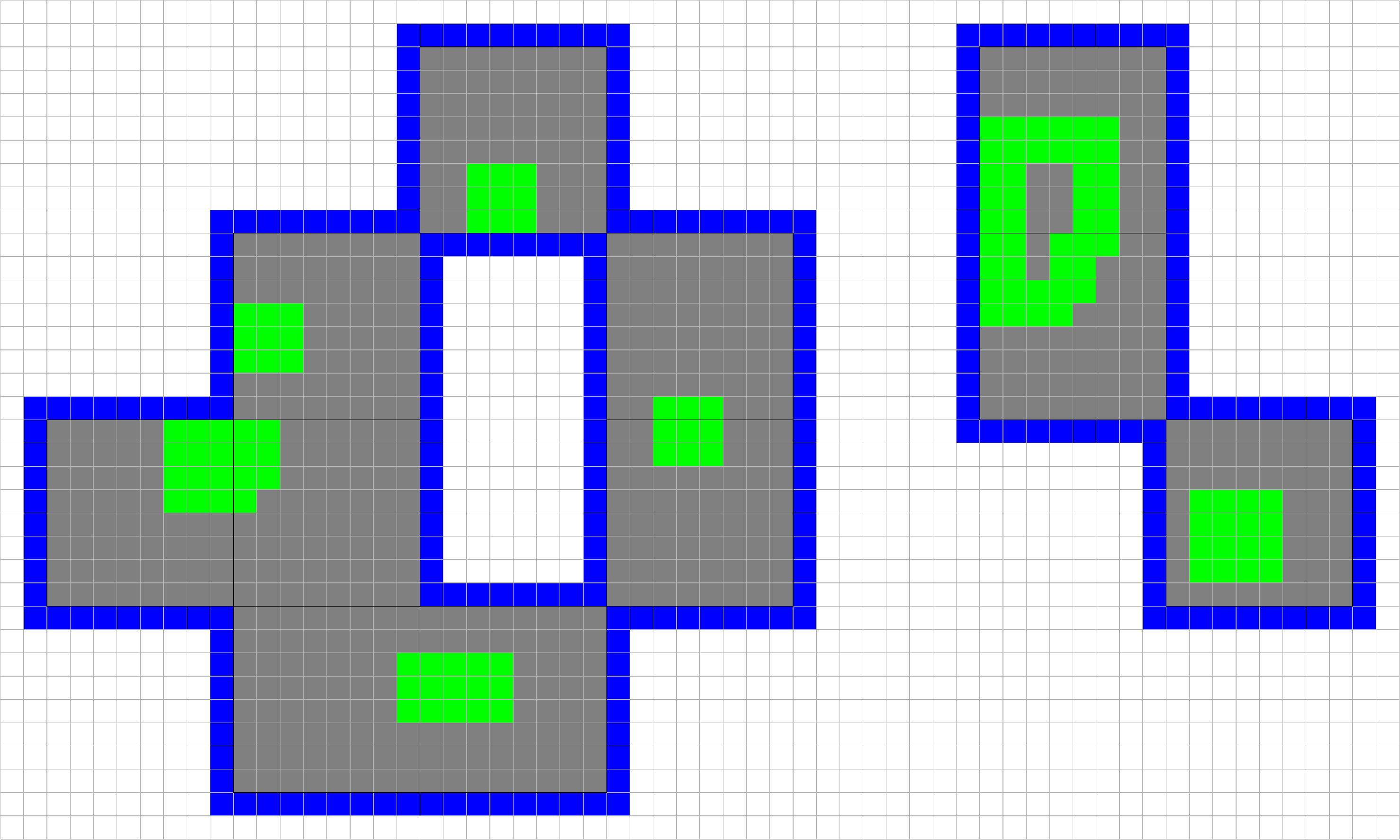}
  \caption{example of a bad region (or, rather, a section of a (3-dimensional) bad region): the green (color online) regions are the bad sampling cubes, the gray or green regions are the bad smoothing cubes, and the blue region consists of the extra cubes added in~(\ref{eqbad}).}
  \label{figure:diagonaledges}
\end{figure}

\begin{definition}[Good and bad regions] \label{def.bad}
Given a spin configuration $\sigma\in\Theta_{\Lambda'}$, a sampling cube $S_\xi$ is said to be

\begin{itemize}
  \item{\bf good} if the spins inside $S_\xi$ are all equal, and $\sigma_\xi\in\{1,2,3\}$. In this case, $\sigma_\xi$ is called the {\it magnetization} of the sampling cube.

  \item{\bf bad} otherwise, that is, every bad sampling cube either contains at least one spin equal to 0 or 4, or it contains at least one pair of neighboring blocks with different spins.
\end{itemize}
Furthermore, we define
\begin{equation}
  B(\sigma):=\bigcup_{\xi \in \Lambda':\ S_\xi\, {\rm is}\, {\rm bad} } S_\xi
\end{equation}
as the union of all bad sampling cubes, as well as the coarser set
\begin{equation}
  B_{s}(\sigma):=\bigcup_{a\in\Lambda'':\  \mathcal S_a\cap B(\sigma)\neq \emptyset} \mathcal S_a\;,
\label{Bcoarser}
\end{equation}
(the lattice $\Lambda''$ and the smoothing cubes $\mathcal S_a$ were defined in section~\ref{section:model}). Finally, we define the ``bad region'' by adding a layer of blocks to $B_s$:
\begin{equation}
  \bar B(\sigma)=\bigcup_{\xi\in\Lambda':\ d'_\infty(\xi,B_s(\sigma))\le1}\Delta_\xi
  \label{eqbad}
\end{equation}
(see  Fig.~\ref{figure:diagonaledges}).
\end{definition}
\bigskip

\noindent{\bf Remark}: In other words, the bad region is a coarse version of the set of blocks which are either empty or contain several plate types, or whose neighbors have a different spin. The reason why we made this set coarser 
is to ensure that plates in different connected components of $\Lambda\setminus\bar B(\sigma)$ do not interact directly, which simplifies the construction of the contour expansion discussed below. 
Indeed, with our choice, different components are at least at an $L_\infty$-distance $2\ell$,
see Fig.\ref{figure:diagonaledges}. 
Moreover, our choice guarantees that the distance between two different connected components of $\bar B(\sigma)$ is larger than $6\ell$, which implies that the effective interaction among contours, called $W^{(\Lambda)}(\partial)$ 
in Lemma \ref{lemma:contours} below, is conveniently small, since it is mediated by at least 
three plates; this condition will be used, in particular, in the proof of Lemma \ref{lemma:polymeractivity}.
\bigskip

Let $\Gamma$ be one of the connected  components of $\bar B(\sigma)$. Let $h_\Gamma+1\ge1$ denote the number of connected components of $\Lambda\setminus\Gamma$. One of these components is adjacent to  $\mathbb Z^3\setminus\Lambda$, and is, naturally, identified as the {\it exterior} of $\Gamma$, which we denote by ${\rm Ext}\,\Gamma$.  When $h_\Gamma\ge1$, the additional connected components of $\Lambda\setminus\Gamma$ are called {\it interiors} of $\Gamma$, which we denote by ${\rm Int}_j\Gamma$, $j=1,\cdots,h_\Gamma$.  
For future reference, we denote by $\mathfrak{Int}$ the set of all possible such interiors, as we let the spin configuration $\sigma$ vary in $\Theta^q_{\Lambda'}$, and the box $\Lambda$ grow. Note that $\Lambda$ is in 
${\mathfrak{Int}}$, and any element $B\in {\mathfrak{Int}}$ satisfies the properties spelled out in the remark after \eqref{Zqdef}. By construction, if $B\in {\mathfrak{Int}}$, then $B$ has no interior. 
\bigskip

\begin{definition}[Contours] \label{def.con}
Given $q\in\{1,2,3\}$, a spin configuration $\sigma\in\Theta_{\Lambda'}^q$ and a plate configuration $P\in\Omega_\Lambda(\sigma)$, we associate a {\it contour} $\gamma:=(\Gamma_\gamma,\sigma_\gamma,P_\gamma)$ 
to each connected component of $\bar B(\sigma)$. Here
\begin{itemize}
  \item $\Gamma_\gamma$ is the connected component of $\bar B(\sigma)$, and is called the {\it support} of the contour;

  \item $\sigma_{\gamma }$ is the restriction of the spin configuration $\sigma$ to $\Gamma_\gamma$;

  \item $P_\gamma$ is the restriction of the plate configuration $P$ to $\Gamma_\gamma$. 
  
  \end{itemize}
  
By the definition of $\bar B(\sigma)$, all the blocks in 
\begin{equation}\label{boh}\partial_{ext}\Gamma_\gamma:=\bigcup_{\xi\in\Gamma'_\gamma:\ d'_\infty(\xi,({\rm Ext}\Gamma_\gamma)')= 1}\Delta_\xi\end{equation} have the same magnetization,
which we denote by $m_{\mathrm{ext},\gamma}\in\{1,2,3\}$. Similarly, all the blocks in
$$\partial_{int,j}\Gamma_\gamma:=\bigcup_{\xi\in\Gamma_\gamma':\ d'_\infty(\xi,({\rm Int}_j\Gamma_\gamma)')= 1}\Delta_\xi$$
have the same magnetization, which we denote by $m_{\mathrm{int},\gamma}^j$. See Fig.\ref{figure:diagonaledges}, where the regions $\partial_{ext}\Gamma_\gamma$ and $\partial_{int,j}\Gamma_\gamma$ 
are colored in blue. 
The collection of all the contours associated with $(\sigma, P)$ is called the {\it contour configuration} associated with $(\sigma, P)$ and will be denoted by the symbol $\partial$. 
\end{definition}
\bigskip

\noindent{\bf Remark:} A contour $\gamma$ must satisfy a number of constraints. For instance: 
\begin{itemize}
\item $\Gamma_\gamma$ must be a union of smoothing cubes, of an external layer and (possibly, if $\cup_j{\rm Int}_j\Gamma_\gamma\neq\emptyset$) 
of internal layers, compatibly with the definition of  $\bar B(\sigma)$, see \eqref{eqbad};
\item the spin configuration $\sigma_\gamma$ must be such that
every sampling cube intersecting $\partial_{ext}\Gamma_\gamma$ is good, with magnetization $m_{\mathrm{ext},\gamma}$, and similarly for the 
sampling cubes intersecting $\partial_{int,j}\Gamma_\gamma$; 
\item the spin configuration $\sigma_\gamma$ must be such that each smoothing cube contained in $\Gamma_\gamma$ intersects at least one bad sampling cube; 
\item $P_\gamma$ must be compatible with $\sigma_\gamma$.
\end{itemize}

We denote the set of {\it possible} contour configurations with $q$ boundary conditions, excluding the empty configuration, by $\mathcal C(\Lambda|q)$: this is the set of non-empty 
contour configurations $\partial$ for which there exist $
\sigma\in\Theta_{\Lambda'}^q$ and $P\in \Omega_\Lambda(\sigma)$ such that $\partial$ is the contour configuration associated to $(\sigma,P)$. The contour configurations $\partial \in \mathcal C(\Lambda|q)$ are fully 
characterized by the following properties: each $\gamma\in\partial$ is possible (that is, there exist $\sigma\in\Theta_{\Lambda'}^q$ and $P\in \Omega_\Lambda(\sigma)$ such that $\gamma$ is one of the contours associated to 
$(\sigma,P)$); $\Gamma_\gamma$ and $\Gamma_{\gamma'}$ are disconnected, for all pairs of distinct contours $\gamma,\gamma'\in\partial$;
the external/internal magnetizations of the contours in $\partial$ satisfy a {\it compatibility condition}, namely, if $\Gamma_\gamma$, with $\gamma\in\partial$, is immediately 
contained\footnote{We say that $\Gamma_\gamma$, with $\gamma\in\partial$, is `immediately contained' in ${\rm Int}_j\Gamma_{\gamma'}$, with $\gamma'\in\partial$, 
if $\Gamma_{\gamma}\subset {\rm Int}_j\Gamma_{\gamma'}$ and there exists no other contour 
$\gamma''\in\partial$ such that $\Gamma_{\gamma}\subset {\rm Int}_{j'}\Gamma_{\gamma''}\subset {\rm Int}_{j}\Gamma_{\gamma'}$.}
in the interior ${\rm Int}_j\Gamma_{\gamma'}$ of another contour $\gamma'\in \partial$, then $m_{\mathrm{ext},\gamma}=m_{\mathrm{int},\gamma'}^j$. 
In terms of these definitions, we can rewrite the partition function~(\ref{Zqdef}) as
\begin{equation}
  Z(\Lambda|q)=Z^q(\Lambda)+\sum_{\partial\in\mathcal C(\Lambda|q)}\mathbf Z_{\partial}(\Lambda|q)
  \label{eqZcontourtmp}
\end{equation}
where $\mathbf Z_{\partial}(\Lambda|q)$ denotes the partition function of plate configurations whose associated contour configuration is $\partial$. Note that the sum over $\mathcal C(\Lambda|q)$ is actually an integral, since it includes an integral over the position of plates inside contour supports. This equality is tautological, and, as such, not all that helpful. Indeed, the compatibility condition among the external/internal magnetizations of the contours, mentioned above,
induces an effective long-range interaction between them, which prevents us from using a cluster expansion to compute the partition function of the contour model. This interaction can be eliminated, as stated in Lemma~\ref{lemma:contours} below.

\subsection{Interacting contour representation}\label{subsection:Zcontour}

In addition to $\mathcal C(\Lambda|q)$, we introduce another set $\mathcal C(\Lambda,q)$ of contour configurations. We say that $\partial \in\mathcal C(\Lambda,q)$, if the following properties are verified:  each contour 
$\gamma\in\partial$ is possible (in the same sense spelled out above, 
see a few lines above \eqref{eqZcontourtmp}); $\Gamma_\gamma$ and $\Gamma_{\gamma'}$ are disconnected, for all pairs of distinct contours $\gamma,\gamma'\in\partial$; 
the external magnetization $m_{\mathrm{ext},\gamma}$ is equal to $q$, for all $\gamma\in\partial$. 
Note that, by definition, the external magnetization of every contour in a contour configuration $\partial\in\mathcal C(\Lambda,q)$ is $q$,
even in situations where a contour is immediately contained in another contour whose internal magnetization is different from $q$. Therefore, the contour configurations
in $\mathcal C(\Lambda,q)$ are not {\it possible} contour configurations in the sense given in the previous section, but this is not a
problem. On the contrary, summing over the contour configurations in $\mathcal C(\Lambda,q)$
is crucial to avoid long-range interactions between contours, thus allowing us to perform a cluster expansion of the contour theory. 
The desired contour representation of the partition function is summarized in the following lemma.

\begin{lemma}[Contour expansion]\label{lemma:contours}
The conditioned partition function $Z(\Lambda|q )$, $q=1,2,3$, can be written as
\begin{equation}
  \frac{Z(\Lambda|q )}{Z^q(\Lambda)}=1+
  \sum_{\partial  \in \mathcal C(\Lambda,q) }   
  \left(\prod_{\gamma \in\partial} \zeta_{q}^{(\Lambda)} (\gamma ) \right) 
  \ e^{W^{(\Lambda)}(\partial)}
  \label{Zcontour}
\end{equation}
where:
\begin{itemize} 
\item $\zeta_{q}^{(\Lambda)}(\gamma)$ is the {\it activity} of $\gamma$:
  \begin{equation}
    \zeta_{q}^{(\Lambda)} (\gamma):=\zeta^0_{q}(\gamma)\,\exp\left(-\int_{\Omega^q_\Lambda} dP\ \varphi^T(P)z^{|P|}F_{\gamma}(P)\right)
    \label{zetadef}
  \end{equation} 
  with
  \begin{equation}
    F_\gamma(P):=
    \left\{\begin{array}{ll}
      1&\mathrm{if\ there\ exist\ two\ plates\ }p_1,\ p_2\ \mathrm{in\ }P\ \mathrm{such\ that}\\
       &p_1\ \mathrm{belongs\ to\ }\Lambda\setminus\Gamma_\gamma\ \mathrm{and\ }p_2\ \mathrm{belongs\ to\ }\Gamma_\gamma,\\[0.3cm]
      1&\mathrm{if\ there\ exists\ a\ plate}\ p_1\ \mathrm{in\ }P\ \mathrm{that\ belongs\ to\ }{\rm Ext}\,\Gamma_\gamma\\
       &\mathrm{and\ a\ plate\ }p_2\ \mathrm{in\ }P_\gamma\ \mathrm{such\ that\ }p_1\cap p_2\neq\emptyset,\\[0.3cm]
      0&\mathrm{otherwise}
    \end{array}\right.
    \label{eqF}
  \end{equation}
  and
  \begin{equation}
    \zeta^0_{q}(\gamma):=\frac{z^{|P_{\gamma}|}{\varphi (P_{\gamma })}}{Z^{q}(\Gamma_\gamma) } \prod_{j=1}^{h_\Gamma}
    \frac{ Z^{(\gamma)} (\mathrm{Int}_{j}\Gamma_\gamma |m^j_{\mathrm{int},\gamma})}{Z(\mathrm{Int}_j\Gamma_\gamma |q )}\;,
    \label{eqz0}
  \end{equation}
  in which $Z^{(\gamma)}(A|m)$, with $A\in\mathfrak{Int}$ (recall that $\mathfrak{Int}$ was introduced right before Definition \ref{def.con}), 
  is the partition function of plates in $A$ with $m$-boundary conditions, with the constraint that plates must not intersect plates in $P_\gamma$, defined in a way analogous to \eqref{Zqdef} (cf. also with the remark after \eqref{Zqdef}): 
\begin{equation} \label{eq:4.10.b}
Z^{(\gamma)}(A|m)=\sum_{\sigma\in\Theta^{m}_{A'}} \int_{\Omega_{A}(\sigma)}
  dP\ \varphi (P\cup P_\gamma)z^{|P|}.\end{equation}
Moreover, the function 
$\varphi^T(P)$ in \eqref{zetadef} is the {\it Ursell function}:
$\varphi^T(\emptyset)=0$ (hence $|P|\ge 1$), $\varphi^T(p)=1$ and, if $n\ge 2$, 
  \begin{equation}
    \varphi^T(p_1,\cdots,p_n):=
    \sum_{\mathfrak g\in\mathcal G^T(n)}\prod_{\{j,j'\}\in\mathcal E(\mathfrak g)}(\varphi(p_j,p_{j'})-1)
  \end{equation}
  in which $\mathcal G^T(n)$ is the set of connected graphs on $n$ labeled vertices, and $\mathcal E(\mathfrak g)$ is the set of undirected edges of the graph $\mathfrak g$. In particular, $\varphi^T(P)$ vanishes if $\bigcup_{p\in P}R_p$ is disconnected.

  \item $W^{(\Lambda)}(\partial)$ is the interaction between the contours in $\partial$: if $|\partial|=1$, then $W^{(\L)}(\partial)=0$, and if $|\partial|\ge2$, then
  \begin{equation}
    W^{(\Lambda)}(\partial) = \int_{\Omega^q_\Lambda}dP\ \varphi^T(P)z^{|P|}
    \sum_{n\ge 2}\frac{(-1)^{n}}{n!}\sum^*_{\gamma_1,\cdots,\gamma_n\in\partial}F_{\gamma_1}(P)\cdots F_{\gamma_n}(P)
  \end{equation}
  where the $*$ on the sum indicates the constraint that $\gamma_1,\cdots,\gamma_n$ are all distinct.
  \end{itemize}
\end{lemma}

\noindent{\bf Remark}:
Note that $F_{\gamma} (P)\neq 0$ only when either $P$ has two intersecting plates, one of which belongs to the contour's support, and the other to its complement (hence $|P|\geq 2$) or $P$ has a plate intersecting 
one of the plates in the contour (in this case we may have $|P|=1$). \medskip

\noindent{\bf Remark}: Note that the second condition in \eqref{eqF} requires that $p_1$ belongs to a block $\Delta_\xi$ such that $d_\infty'(\xi,\Gamma_\gamma)\le2$. 
\medskip

\noindent{\bf Remark}: As we will prove in the following, the {\it interaction} $e^{W^{(\Lambda)}(\partial)}$ is a short-range interaction, that is, it decays exponentially with the distance between contours. This property is essential to the convergence of the cluster expansion of the contour model.
\medskip

\noindent{\bf Remark}: For future reference, we note that the constrained partition function in \eqref{eq:4.10.b} can be rewritten in a form that does not involve summation over spins:
\begin{equation}Z^{(\gamma)}(A|m)=\int_{\Omega^m_{\partial A}} 
  dP\ \varphi (P\cup P_\gamma)z^{|P|}\int_{\Omega_{A^{\circ}}} 
  d\tilde P\ \varphi (\tilde P\cup P)z^{|\tilde P|},\end{equation}
where $\partial A=\cup_{\xi\in A': d_\infty'(\xi,({\rm Ext}A)')\le 8}\Delta_\xi$ is the layer of blocks that are uniformly magnetized by the boundary conditions, and $A^\circ=A\setminus \partial A$. On the other hand, 
this expression is equivalent to 
\begin{equation}Z^{(\gamma)}(A|m)=\int_{\Omega^m_{\partial A}\setminus V_m(P_\gamma)} 
  dP\ \varphi (P)z^{|P|}\int_{\Omega_{A^{\circ}}} 
  d\tilde P\ \varphi (\tilde P\cup P)z^{|\tilde P|}=: Z(A\setminus V_m(P_\gamma)\,|\,m),\end{equation}
where $V_m(P_\gamma)$ is the excluded volume created by the plates in $P_\gamma$ on those in $P$. Note that $A\setminus V_m(P_\gamma)$ is an element of $\mathfrak{Int}'$, where
\begin{equation} \mathfrak{Int}':=\Bigg\{A\setminus V:\ A\in\mathfrak{Int}\ {\rm and}\ V\subset \mathbb R^3\ {\rm such}\ {\rm that}\ V\subset\hskip-.5truecm \bigcup\limits_{\substack{\xi\in A':\\ d_\infty'(\xi,({\rm Ext}A)')\le 2}}
\hskip-.5truecm\Delta_\xi\Bigg\}.\label{int'}\end{equation}
\medskip

\indent The idea of the proof is to first map the plate model to one of {\it external} contours, which are contour configurations such that a contour may not lie inside another. We then rewrite the partition function as a sum over external contours of the activity of the contour times the partition function {\it inside} each contour. Now, the boundary of this partition function is dictated by the internal magnetizations of the contours. To remove this dependence, we replace the boundary condition with $q$, at the price of including an extra factor in the activity of the contour, which is the second ratio in~(\ref{eqz0}). The construction is then iterated, and yields a model of contours whose external magnetization is {\it always} $q$. This eliminates the long-range interaction between contours. The short-range interaction, mediated by the plates between contours, which are, by construction, all of type $q$, is then computed using a Mayer expansion.
\bigskip

\indent The proof of this lemma is entirely analogous to that of~\cite[Lemma~1]{DG13}, and is left to the reader.

\subsection{Hard core polymer representation}\label{subsection:polymer}

The contours in~(\ref{Zcontour}) interact with each other, due to the presence of the many-body potential $W^{(\Lambda)}(\partial)$. In order to set up the cluster expansion, we will first map the interacting contour model to a hard core polymer model. In order to formulate our next technical lemma, we need a couple more definitions. We let $\mathfrak B(\Lambda)$ be the set of unions of blocks in $\Lambda$, and 
$\mathfrak B^T(\Lambda)$ the set of $D$-connected unions of blocks in $\Lambda$ (with
the prefix ``$D$''  meaning ``diagonal''): here we say that two blocks are $D$-connected if they touch either on a face, or on an edge or at a corner; of course, if they are not $D$-connected, we say that they 
are $D$-disconnected. 

\bigskip

\begin{lemma}[Polymer expansion]\label{lemma:polymer} We have
\begin{equation}
  \frac{Z(\Lambda|q)}{Z^q(\Lambda)}=1+\sum_{n\ge 1}\frac1{n!}\sum_{X_1,\ldots,X_n\in \mathfrak B^T(\Lambda)}\phi(X_1,\ldots,X_n)\prod_{i=1}^nK_q^{(\Lambda)}(X_i)
  \label{eqpolymerZ}
\end{equation}
where:
\begin{itemize}
  \item  $\phi(\{X_1,\cdots,X_m\})\in\{0,1\}$ is equal to 1 if and only if $X_i$ and $X_j$ are $D$-disconnected for all $i\neq j$.
  \item $K_q^{(\Lambda)}(X)$ is the {\it activity} of $X$:
  \begin{equation}
    K_{q}^{(\Lambda)}(X):=K_{q,1}^{(\Lambda)}(X)+K_{q,\ge2}^{(\Lambda)}(X)\label{k1k2}
  \end{equation}
  with
  \begin{equation}
    K_{q,1}^{(\Lambda)}(X):=
    \sum_{\mAthop{\gamma\in\mathcal C_1(\Lambda,q)}_{\Gamma_\gamma=X}}\zeta_{q}^{(\Lambda)}(\gamma)
    \label{K1}
  \end{equation}
  and
  \begin{equation}
    K_{q,\ge2}^{(\Lambda)}(X):=
    \kern-10pt
    \sum_{\mAthop{X_0,X_1\in\mathfrak B(X)}_{X_0\cup X_1=X}}
    \sum_{\mAthop{\partial\in\mathcal C_{\ge2}(\Lambda,q)}_{\Gamma_\partial=X_0}}
    \left(\prod_{\gamma\in\partial}\zeta_{q}^{(\Lambda)}(\gamma)\right)
    \sum_{p\ge1}{\frac{1}{p!}}\sum_{\mAthop{Y_1,\cdots,Y_p\subset\mathfrak B^T(X)}_{Y_1\cup\cdots\cup Y_p=X_1}}^*
    \left(\prod_{j=1}^p\left(e^{\mathcal F_\partial(Y_j)}-1\right)\right)
    \label{Kdef}
  \end{equation}
in which: $\mathcal C_1(\Lambda,q)$ and $\mathcal C_{\ge2}(\Lambda,q)$ denote the sets of contour configurations with, respectively, a single contour, and at least two contours;
$\Gamma_\partial\equiv\bigcup_{\gamma\in\partial}\Gamma_\gamma$; the $*$ on the sum indicates that the sets $Y_1,\cdots,Y_p$ are different from each other; 
  \begin{equation}
    \mathcal F_\partial(Y):=\sum_{n\ge2}{\frac{(-1)^{n}}{n!}}
    \sum_{\gamma_1,\cdots,\gamma_n\subset\partial}^* \int_{\Omega^q_\Lambda}dP\ z^{|P|}\varphi^T(P)F_{\gamma_1}(P)\cdots F_{\gamma_n}(P)\mathfrak I_Y(P)
    \label{Fdef}
  \end{equation} 
  where  $\mathfrak I_Y(P)\in\{0,1\}$ is equal to 1 if and only if $Y$ is the smallest $D$-connected union of blocks that is such that every plate is {\it contained} in $Y$ (that is, the support of every plate is a subset of $Y$).
\end{itemize}
\end{lemma}

\noindent{\bf Remark:} The sets $Y_1,\cdots,Y_p$ are not necessarily disconnected, but they are different
from each other. By the definitions of $\mathcal F_\partial(Y)$ and $F_\gamma(P)$, it follows that $\mathcal F_\partial(Y)\neq 0$ (that is, $e^{\mathcal F_\partial(Y_j)}-1\neq 0$) 
only if $Y$ is $D$-connected with at least two contours in $\partial$: in order to prove this fact, it is crucial that every plate is {\it contained} in $Y$. 
Therefore, the sum over $Y_1,\ldots,Y_p$ in \eqref{Fdef} can be restricted to elements in $\mathfrak B^T(\Lambda)$ that 
are $D$-connected with at least two contours in $\partial$. 
\bigskip

\indent The proof of this lemma is fairly straightforward, and virtually identical to~\cite[Lemma~2]{DG13}.
The key identity is
\[
e^{W^{(\Lambda )} (\partial)} = e^{\sum_{n\geq 2} \frac{(-1)^{n}}{n!} \sum^*_{\gamma_1,\cdots,\gamma_n\in\partial} 
\int_{\Omega^q_\Lambda}dP\ \varphi^T(P)z^{|P|}F_{\gamma_1}(P)\cdots F_{\gamma_n}(P)}=
\prod_{Y\in \mathfrak B^{T}(\Lambda)} \left[ (e^{ \mathcal F_\partial(Y)} -1)+1\right].
\]
The only real difference is that the sets $Y_i$ cover all the plates responsible for the interaction between contours, whereas in~\cite{DG13}, only the extremal blocks are kept (in~\cite{DG13}, the analog of the sets $Y_i$ are 
denoted by $\overline Y_i$). The details are left to the reader. 

\section{Convergent cluster expansion}\label{section:polymer_cluster}

In this section we prove the convergence of the contour expansions introduced above. The results of this section justify a posteriori the 
definitions given in the previous section, in particular the specific form of the contour representation and of the polymer expansion that we chose and introduced. 
The key problem is to estimate the contour and polymer activities, which is not trivial, since they involve ratios of partition functions in their interiors, see \eqref{eqz0}, 
which must be estimated inductively. 
Once a smallness condition 
on the activities is known, the convergence of the expansion is `trivial', in the sense that it follows from the classical theory of the cluster expansion. 
The main convergence result of this section is summarized in the following theorem. 

\begin{theorem}[Polymer cluster expansion]\label{theorem:cluster}
  Given $\alpha\in(\frac34,1]$, if $zk^{3-\alpha}$ and $\log k/(zk^{3\alpha})$ are sufficiently small, then 
  \begin{equation}
    |K_q^{(\Lambda)}(X)|\le \bar\epsilon^{|X'|}
    \label{eqbK}
  \end{equation}
  with
  \begin{equation}\label{epsbar}
    \bar\epsilon:=\max\{(zk^2)^{\cst c{c.1}}, e^{-\cst c{c.2} zk^{2+\alpha}}\}, 
  \end{equation}
for suitable constants $\cst c{c.1}, \cst c{c.2}>0$. Furthermore,
  \begin{equation}
    \log Z(\Lambda|q)=\log Z^q(\Lambda)+\sum_{n\ge 1}\frac1{n!}\sum_{X_1,\ldots, X_n\in\mathfrak B^T(\Lambda)}\phi^T(X_1,\ldots, X_n)\prod_{i=1}^nK_q^{(\Lambda)}(X_i)
    \label{eqce}
  \end{equation}
  where $\phi^T$ is the {\it Ursell function}: $\phi^T(\emptyset)=0$, $\phi^T(X)=1$ and, if $n\ge 2$, 
  \begin{equation}
    \phi^T(X_1,\cdots,X_n):=
    \sum_{\mathfrak g\in\mathcal G^T(n)}\prod_{\{j,j'\}\in\mathcal E(\mathfrak g)}(\phi(X_j,X_{j'})-1)
  \end{equation}
  in which $\mathcal G^T(n)$ is the set of connected graphs on $n$ labeled vertices, and $\mathcal E(\mathfrak g)$ is the set of undirected edges of the graph $\mathfrak g$. In particular, $\phi^T(X_1,\ldots, X_n)$ vanishes if $\bigcup_{i}X_i$ is $D$-disconnected. Finally,  the sum in the right side of~(\ref{eqce}) is absolutely convergent: for all $X_0\in\mathfrak B^T(\Lambda)$, $\forall m\ge1$,
  \begin{equation}
\sum_{n\ge m}\frac1{n!}    \sum_{X_1,\ldots, X_n\in\mathfrak B^T(\Lambda)}\left|\phi^T(X_0,X_1,\ldots,X_n)\prod_{i=1}^nK_q^{(\Lambda)}(X_i) \right|
    \le C |X_0'|\bar\epsilon^{500\, m} ,
    \label{eqcvce}
  \end{equation}
for a suitable constant $C>0$, where $500=10^3/2$ and $(10\ell)^3$ is the value of $|X|$ for the smallest possible contour of non-vanishing activity. 
\end{theorem}
  
\indent Here we will focus on~(\ref{eqbK}), since the rest of the theorem follows from the general theory of cluster expansions for polymer models, which is standard (see, for instance, \cite{Ru99,Br86,KP86,BZ00,GBG04}). We will proceed in two steps.
\begin{itemize}
  \item The first is to prove that, provided the activity $\zeta_q^{(\Lambda)}(\gamma)$ of a contour $\gamma$, defined in~(\ref{zetadef})
and \eqref{eqz0}, decays as $e^{-(\mathrm{const.})\ zk^{2+\alpha}|\Gamma_\gamma'|}$, then \eqref{eqbK} holds. 
This follows from the fact that the factor $e^{\mathcal F_{\partial}(Y_j)}-1$ appearing in~(\ref{Kdef}) is exponentially small in the size of $Y_j$, or, in other words, that the interaction between contours is of short range.

  \item The second step is to prove that $\zeta_q^{(\Lambda)}(\gamma)$ is bounded by $e^{-(\mathrm{const}.)\ zk^{2+\alpha}|\Gamma_\gamma'|}$. The basic idea of the proof is that 
the number of bad blocks or dipoles in $\Gamma_\gamma$ is proportional to its rescaled volume $|\Gamma_\gamma'|$,
and the weight of a bad block or dipole is, by Lemma~\ref{lemma:twodircubes} and Corollary~\ref{corollary:twodircubes}, 
bounded by $e^{-(\mathrm{const}.)\ zk^{2+\alpha}}$. A complication comes from the fact that the bound on $\zeta_q^{(\Lambda)}(\gamma)$ 
requires an inductive argument to estimate the 
ratio of partition functions in~(\ref{eqz0}): for this purpose, we use Theorem~\ref{theorem:cluster} inductively, starting from the contours that are so small that their interior cannot contain 
other contours, and then moving to larger and larger contours. 
\end{itemize}

\subsection{Polymer activity}
Here we discuss the first step anticipated above: namely, we assume the desired bound on the contour activity, and, on the basis of this hypothesis, we deduce bounds on the polymer activity. 
From now on, $C,C', \ldots$, and $c,c',\ldots$, indicate universal
positive constants (to be thought of as ``big'' and ``small'', respectively), whose specific
values may change from line to line.
\bigskip

\begin{lemma}[Polymer activity]\label{lemma:polymeractivity}
If $zk^{2}$ and $1/(zk^{2+\alpha})$ are sufficiently small and, for every contour $\gamma$,
\begin{equation}
  \left|\int_{\Omega_{\Gamma_\gamma}(\sigma_\gamma)}dP_\gamma\ \zeta_{q}^{(\Lambda)}(\gamma)\right|\le e^{-\cst c{c.3} zk^{2+\alpha}|\Gamma_\gamma'|}
  \label{eqconditionconvergence}
\end{equation}
for some constant $\cst c{c.3}>0$, then the polymer activity satisfies \eqref{eqbK}, that is, 
\begin{equation}
  |K_q^{(\Lambda)}(X)|\le \bar\epsilon^{|X'|}
  \label{eqboundK}
\end{equation}
where $\bar\epsilon$ was defined in \eqref{epsbar}. 
\end{lemma}
\bigskip

{\bf Proof of lemma~\ref{lemma:polymeractivity}.} The main idea of the proof is to extract from
$(e^{\mathcal F_\partial(Y)}-1)$ an exponential decay proportional to $(zk^2)^{c|Y'|}$. 
This is due to the fact that the only plate configurations $P$ contributing to \eqref{Fdef} are the connected ones 
(here we say that two plates $p,p'$ are connected if $p\cap p'\neq\emptyset$): therefore, the number of plates in $P$ 
must be at least proportional to $|Y'|$. Since, as we will show below, every additional plate after the first one in $P$ comes with a factor $zk^2$, 
we find that $\mathcal F_\partial(Y)$ decays like $(zk^2)^{c|Y'|}$, and similarly for $(e^{\mathcal F_\partial(Y)}-1)$. 
After having extracted this exponential decay, we can insert the bound 
on $\zeta$ as in~(\ref{eqconditionconvergence}) and perform the sum over $X_{0},X_{1}.$
\bigskip

\indent Recalling \eqref{k1k2} and \eqref{K1}, we first bound 
\begin{equation}
  |K_{q,1}^{(\Lambda)}(X)|\le5^{|X'|}e^{-\cst c{c.3} zk^{2+\alpha}|X'|}
  \label{ineqK1}
\end{equation}
in which the factor $5$ comes from enumerating the spin configurations $\sigma_\gamma$ in the contour and we used \eqref{eqconditionconvergence} for integrating over the plate configurations at fixed $\sigma_\gamma$. This implies the analog of (\ref{eqboundK}) for $K_{q,1}^{(\Lambda)}(X)$.
\bigskip

\indent The key ingredient in the rest of the proof is the Mayer expansion of the plate model. Once again, we will not discuss this
expansion in detail, as it follows from the general theory of cluster expansions~\cite{Ru99,Br86,KP86,BZ00,GBG04}.
Recalling the definitions of $\Omega^q_S$ and $\Omega^{n,q}_S$ given right before \eqref{mayer0.1}, we let $\Omega^{\ge l,q}_S=\cup_{n\ge l}\Omega^{n,q}_S$
be the set of plate configurations of type $q$ with center in $S$ and at least $l$ plates. Using a Mayer expansion it can be proved that, for any $S\subset \mathbb R^3$,
\begin{equation}
  \int_{\Omega_\Lambda^{\ge l,q}} dP\ z^{|P|}|\varphi^T(P)|\mathds{1}(p_1\ {\rm belongs}\ {\rm to}\ S)   \le C^l z|S| (zk^2)^{l-1}  \label{eqcvce_plate}
\end{equation}
for some constant $C>0$, where $p_1$ is the first plate in $P$ (recall that the integration measure is symmetric under permutations of the plates in $P$). 
We now want to use this estimate to bound \eqref{Kdef}. The key point is to estimate the sum over $p$ in the right side of \eqref{Kdef}. We claim that
\begin{equation}\label{eq:psum}
\sum_{p\ge1}\frac{1}{p!}\sum_{\mAthop{Y_1,\cdots,Y_p\subset\mathfrak B^T(X)}_{Y_1\cup\cdots\cup Y_p=X_1}}^*
    \left(\prod_{j=1}^p\left |e^{\mathcal F_\partial(Y_j)}-1\right |\right)\leq
      (zk^2)^{\frac{c_0}2|X_1'|} e^{|X_{0}'|zk^{3} O (zk^{2})  }.
\end{equation}
To prove this bound we  use (\ref{eqcvce_plate}) to
estimate $\mathcal F_\partial(Y)$. Note that the $L_\infty$ distance between the centers of two overlapping plates is,
at most, $k\equiv2\ell$. Since the distance between two disconnected contours is, at least, $6\ell$,  and any plate configuration 
$P$ contributing to $\mathcal F_\partial(Y)$ must intersect plates belonging to the supports of at least two disconnected contours (due to the constraints induced by the 
functions $F_{\gamma_i}$), then any plate configuration $P$ contributing to $\mathcal F_\partial(Y)$ must contain at least 3 plates. Moreover, by a similar argument, 
it must contain at least $1+c_0|Y'|$ plates, for a suitable constant $c_0$, which can be chosen, e.g., to be $1/14$. Note also that $\mathcal F_\partial(Y)$ is non zero only if $\mathrm{dist}(Y,X_0)=0$, where 
dist is the Euclidean distance. 
Therefore, letting: $l_Y:=1+\max(2,c_0|Y'|)$, $N$ be the number of contours in $\partial$ that are $D$-connected to the set $Y$,
$\D_{\xi_1}$ be the `first' block of $Y$ (with respect to any given order of its blocks) and $S_Y$ the union of the sampling cubes intersecting $\D_{\xi_1}$, 
\begin{align}
 |\mathcal F_\partial(Y)|&\le 
 \sum_{n=2}^{N}\frac{1}{n!}\sum_{\gamma_1,\cdots,\gamma_n\subset\partial}^*\int_{\Omega^{\ge l_Y,q}_\Lambda}dP\ z^{|P|}|\varphi^T(P)|\mathds{1}(p_1\ {\rm belongs}\ {\rm to}\ S_Y)
 \mathds 1_{\mathrm{dist}(Y,X_0)=0}
\label{boundeF}\\
& \leq 2^{N} zk^{3} (Czk^2)^{\max(2,c_0|Y'|)}\mathds 1_{\mathrm{dist}(Y,X_0)=0} \leq 
zk^3(C'zk^2)^{\max(2, c_0|Y'|)}\mathds 1_{\mathrm{dist}(Y,X_0)=0}\nonumber
\end{align}
for some constants $C,C'>0$, where we used \eqref{eqcvce_plate} and, in the final bound, we used $N\leq |Y'|$.
Moreover we have
\[
\prod_{j=1}^p\left |e^{\mathcal F_\partial(Y_j)}-1\right | \leq e^{\sum_{j=1}^{p}  |\mathcal F_\partial(Y_j) | } \prod_{j=1}^p |\mathcal F_\partial(Y_j) |,
%
\]
where
\[
\sum_{j=1}^{p}  |\mathcal F_\partial(Y_j) |\leq \sum_{\mAthop{Y\in\mathfrak B^{T}(X_{1})}_{\mathrm{dist}(Y,X_0)=0}} |\mathcal F_\partial(Y) |
\leq  C'' zk^3(zk^2)^2|X_0'|
\]
and, using $\sum_{j}|Y_{j}'|\geq |X_{1}'|,$
\[
\prod_{j=1}^p |\mathcal F_\partial(Y_j) |\leq  (zk^2)^{\frac{c_0}2|X_1'|}
\prod_{j=1}^p  |\mathcal F_\partial(Y_j) | (zk^2)^{-\frac{1}{2}\max(2,c_0|Y'_{j}|)}
\]
Inserting these estimates in the sum over $p$
\begin{multline*}
\sum_{p\ge1}\frac{1}{p!}\sum_{\mAthop{Y_1,\cdots,Y_p\subset\mathfrak B^T(X)}_{Y_1\cup\cdots\cup Y_p=X_1}}^* \prod_{j=1}^p  |\mathcal F_\partial(Y_j) | (zk^2)^{-\frac{1}{2}\max(2,c_0|Y'_{j}|)} \\
\leq \sum_{p\ge1}\frac{1}{p!} \Biggl(
      \sum_{\displaystyle\mathop{\scriptstyle Y\in\mathfrak B^T(X)}_{\mathrm{dist}(Y,X_0)=0}}
      |\mathcal F_\partial(Y) | (zk^2)^{-\frac{1}{2}\max(2,c_0|Y'|)}
    \Biggr)^p\\
\le \exp \Biggl(
      \sum_{\displaystyle\mathop{\scriptstyle Y\in\mathfrak B^T(X)}_{\mathrm{dist}(Y,X_0)=0}}
      |\mathcal F_\partial(Y) | (zk^2)^{-\frac{1}{2}\max(2,c_0|Y'|)}
    \Biggr) \leq e^{zk^3|X_0'|O(zk^2)}.
\end{multline*}
Putting the terms together we get \eqref{eq:psum}.
Finally,  inserting the bound (\ref{eqconditionconvergence}) on $\zeta$,
\begin{equation}
  |K_{q,\ge2}^{(\Lambda)}(X)|\leq
  \sum_{\mAthop{X_0,X_1\in\mathfrak B(X)}_{X_0\cup X_1=X,\ X_0\neq\emptyset}}
  5^{|X_0'|}e^{-\cst c{c.3} zk^{2+\alpha}(1+O(zk^{3-\alpha}))|X'_0|}
  (zk^2)^{\frac{c_0}2|X_1'|}
\end{equation}
where we used $zk^{3}zk^{2}=zk^{2+\alpha } zk^{3-\alpha}.$ This  yields~(\ref{eqboundK}).
\qed

\subsection{The activity of contours}\label{subsection:contour_activity}

We will now prove that~(\ref{eqconditionconvergence}) holds, which proves the convergence of the cluster expansion, and concludes the proof of~(\ref{eqbK}).
\bigskip

\begin{lemma}[Contour activity]\label{lemma:activities}
If $zk^{3-\alpha}$ and $\log k/(zk^{3\alpha})$ are sufficiently small, then
\begin{equation}
  \int_{\Omega_{\Gamma_\gamma}(\sigma_\gamma)}dP\ |\zeta_{q}^{(\Lambda)}(\gamma)|\le e^{-\cst c{c.3} zk^{2+\alpha}|\Gamma_\gamma'|}
  \label{boundzeta}
\end{equation}
where $\cst c{c.3}$ is the same constant appearing in \eqref{eqconditionconvergence}.
\end{lemma}

Recall (see~(\ref{zetadef})) that
\begin{equation}
  \begin{largearray}
    \int_{\Omega_{\Gamma_\gamma}(\sigma_\gamma)} dP_\gamma\ \zeta_{q}^{(\Lambda)}(\gamma)=
    \int_{\Omega_{\Gamma_\gamma}(\sigma_\gamma)} dP_\gamma\ 
    \left(\frac{z^{|P_\gamma|}\varphi(P_\gamma)}{Z^q(\Gamma_\gamma)}\right)
    \left(\prod_{j=1}^{h_{\Gamma_\gamma}}\frac{Z^{(\gamma)}(\mathrm{Int}_j\Gamma_\gamma|m_{\mathrm{int},\gamma}^j)}{Z(\mathrm{Int}_j\Gamma_\gamma|q)}\right)
    \cdot\\[0.5cm]\hfill\cdot
    \exp\left(-\int_{\Omega_\Lambda^q}dP\ \varphi^T(P)z^{|P|}F_\gamma(P)\right).
  \end{largearray}
  \label{eqineqzeta2}
\end{equation}
In order to prove Lemma~\ref{lemma:activities}, we bound each factor in~(\ref{eqineqzeta2}), which is done in Lemma~\ref{lemma:activities_3}, \ref{lemma:activities_plates} and~\ref{lemma:activities_ratio}, stated below.
\bigskip

\begin{lemma}\label{lemma:activities_3}
If $zk^{2}$ is sufficiently small, then
\begin{equation}
  \exp\left(-\int_{\Omega_\Lambda^q}dP\ \varphi^T(P)z^{|P|}F_\gamma(P)\right)
  \le e^{O(zk^3zk^2)|\Gamma_\gamma'|}
  \label{eqineqzeta3}
\end{equation}
\end{lemma}

\begin{lemma}\label{lemma:activities_plates}
If $zk^{3-\alpha}$ and $\log k/(zk^{3\alpha})$ are sufficiently small, then
\begin{equation}
  \int_{\Omega_{\Gamma_\gamma}(\sigma_\gamma)} dP_\gamma\ 
  \frac{z^{|P_\gamma|}\varphi(P_\gamma)}{Z^q(\Gamma_\gamma)}
  \le e^{-c zk^{2+\alpha}|\Gamma_\gamma'|}
  \label{eqineqzetam}
\end{equation}
for some constant $c>0$.
\end{lemma}

\begin{lemma}\label{lemma:activities_ratio}
If $zk^{3-\alpha}$ and $\log k/(zk^{3\alpha})$ are sufficiently small, then, if $A\in\mathfrak{Int}$ (recall that $\mathfrak{Int}$ was introduced right before Definition \ref{def.con}) and $m,q\in\{1,2,3\}$,
\begin{equation}
  \frac{Z(A|m)}{Z(A|q)}\le
  e^{c(zk^3zk^2+\bar\epsilon^C)|(\partial_{ext}A)'|}
  \label{eqineqzetas}
\end{equation}
for some constants $c,C>0$, where $\bar\epsilon$ was defined in \eqref{epsbar} and $\partial_{ext}A$ is defined in the same way as \eqref{boh}. 
\end{lemma}
\bigskip

{\bf Remark}: The constrained partition function $Z^{(\gamma)}(\mathrm{Int}_j\Gamma_\gamma|m_{\mathrm{int},\gamma}^j)$ appearing in the right side of 
\eqref{eqineqzeta2} is smaller than the unconstrained partition function $Z(\mathrm{Int}_j\Gamma_\gamma|m_{\mathrm{int},\gamma}^j)$. Therefore, 
Lemma \ref{lemma:activities_ratio} is enough for bounding the ratio $\prod_{j=1}^{h_{\Gamma_\gamma}}\frac{Z^{(\gamma)}(\mathrm{Int}_j\Gamma_\gamma|m_{\mathrm{int},\gamma}^j)}{Z(\mathrm{Int}_j\Gamma_\gamma|q)}$ 
in \eqref{eqineqzeta2}. Combining this remark with Lemma~\ref{lemma:activities_3}, \ref{lemma:activities_plates} and~\ref{lemma:activities_ratio}, we obtain Lemma~\ref{lemma:activities}. 
\bigskip

{\bf Proof of Lemma~\ref{lemma:activities_3}.} The main idea of the proof is to use the Mayer expansion of the plate model
 to extract a dominating term, which is negative, and bound the remainder.
\bigskip

We split
\begin{equation}
  -\int_{\Omega_\Lambda^q}dP\ \varphi^T(P)z^{|P|}F_\gamma(P)
  =-\int_{\Omega_\Lambda^{1,q}} dp\ zF_\gamma(\{p\})
  -\int_{\Omega_\Lambda^{\ge 2,q}}dP\ \varphi^T(P)z^{|P|}F_\gamma(P)
\end{equation}
where we recall that $\Omega_\Lambda^{1,q}$ (resp. $\Omega_\Lambda^{\ge 2,q}$)
is the set of plate configurations of  type $q$ with 1 plate (resp. at least 2 plates).
The first term is non-positive, and the second is bounded by 
\begin{equation}
\Big|  \int_{\Omega_\Lambda^{\ge 2,q}}dP\ \varphi^T(P)z^{|P|}F_\gamma(P)\Big|\le 
\int_{\Omega_\Lambda^{\ge 2,q}}dP\ |\varphi^T(P)|z^{|P|}\mathds 1(p_1\ {\rm belongs}\ {\rm to}\ S_\gamma),\end{equation}
where $S_\gamma=\cup_{\xi: d_\infty'(\xi,\Gamma_\gamma')\le 2}\Delta_\xi$ (here we used the second remark after Lemma \ref{lemma:contours}). We have $|S_\gamma|\le 2|\Gamma_\gamma|$. We are now in the position of applying \eqref{eqcvce_plate}, which gives
\begin{equation}\Big|  \int_{\Omega_\Lambda^{\ge 2,q}}dP\ \varphi^T(P)z^{|P|}F_\gamma(P)\Big|
\le O(zk^3zk^2|\Gamma_\gamma'|).
\end{equation}
\qed
\bigskip

{\bf Proof of Lemma~\ref{lemma:activities_plates}.} 
Let $\sigma_\gamma$ be a spin configuration compatible with the 
fact that $\gamma$ is a contour. As a consequence, every smoothing cube contained in $\Gamma_\gamma$ has
non zero intersection with at least one bad sampling cube; moreover, by its very definition, each
such bad cube must contain either one block with magnetization equal to $0$ or $4$, or one pair of
neighboring blocks with magnetization $q,q'\in\{1,2,3\}$ such that $q\neq q'$ (a `bad dipole' in the sense of Corollary \ref{corollary:twodircubes}). 
Therefore, given $\sigma_\gamma$, it is 
possible to exhibit a partition $\mathcal P$ of $\Gamma_\gamma$ such that: (i) all the elements of the partition consist either 
of a single block or of a bad dipole; (ii) if $\mathfrak M_\gamma$ is the set of blocks in $\mathcal P$ with magnetization 
equal to 0 or $4$ and $\mathfrak D_\gamma$ is the set of bad dipoles in $\mathcal P$, then $|\mathfrak M_\gamma|+|\mathfrak D_\gamma|\ge c'|\Gamma_\gamma'|$, for a constant $c'$ that can be chosen, e.g., equal to $8^{-4}$.
We also let $\mathfrak N_\gamma$ be the set of blocks in $\mathcal P$ that are not in $\mathfrak M_\gamma\cup \mathfrak D_\gamma$. We then bound
\begin{equation}
  \begin{largearray}
    \int_{\Omega_{\Gamma_\gamma}(\sigma_\gamma)} dP_\gamma\ 
    \frac{z^{|P_\gamma|}\varphi(P_\gamma)}{Z^q(\Gamma_\gamma)}
    \\[0.5cm]\hfill
    \le
    \left(\prod_{\mu\in\mathfrak M_\gamma}\frac{2+Z_{\ge2}(\mu)}{Z^q(\mu)}\right)
    \left(\prod_{\delta\in\mathfrak D_\gamma}\frac{Z_{\ge2}(\delta)}{Z^q(\delta)}\right)
    \left(\prod_{n\in\mathfrak N_\gamma}\frac{Z^{\sigma_n}(n)}{Z^q(n)}\right)
    e^{O(zk^3zk^2)|\Gamma'_\gamma|}
  \end{largearray}
\end{equation}
where the $2$ in  the first factor in the right side is due to the activity associated with spin 0, see \eqref{eq:2.8}, and the factor $e^{O(zk^3zk^2)|\Gamma'_\gamma|}$ comes from splitting $Z^q$ into blocks and dipoles, as per~(\ref{mayer1}). We now use Lemma~\ref{lemma:twodircubes} and Corollary~\ref{corollary:twodircubes}, 
and note that $Z^{\sigma_n}(n)=Z^q(n)$, thus getting 
\begin{equation}
  \int_{\Omega_{\Gamma_\gamma}(\sigma_\gamma)} dP_\gamma\ 
  \frac{z^{|P_\gamma|}\varphi(P_\gamma)}{Z^q(\Gamma_\gamma)}
  \le e^{- c'' zk^{2+\alpha}(|\mathfrak M_\gamma|+|\mathfrak D_\gamma|)}
  e^{O(zk^{2+\alpha}zk^{3-\alpha})|\Gamma'_\gamma|}
\end{equation}
for some constant $c''>0$. The result follows from $|\mathfrak M_\gamma|+|\mathfrak D_\gamma|\ge c'|\Gamma_\gamma'|$. 
\qed
\bigskip

{\bf Sketch of the proof of Lemma~\ref{lemma:activities_ratio}.} The main idea of the proof is the following. If $A$ did not contain any contours, it would only contain a single type of plates, and we would be able to express its partition function using a convergent Mayer expansion, and find that the ratio of partition functions only involves clusters that straddle the boundary of $A$. This gives us the appropriate bound, since clusters with at least two plates contribute a weight $zk^3zk^2$. When $A$ contains contours, we proceed by induction and use the fact that, by the inductive hypothesis, the polymer theory inside $A$ admits a convergent cluster expansion. We then show that the only polymer clusters that contribute to the ratio of partition functions are those that straddle the boundary.
\bigskip

\indent The details of the proof are in direct analogy with the proof of~\cite[Lemma~5]{DG13}, and are left to the reader.

\section{Nematic order}\label{section:nematic}

\indent In this section, we give the proof of Theorem~\ref{theorem:nematic}, which follows from a simple modification of the cluster expansion in Theorem~\ref{theorem:cluster}.
We recall that in order to compute density correlations, we need to promote the activity to be plate-dependent, that is, it is a function $\tilde z(p)$. The expansions described in the previous sections 
hold also in this case with the natural modifications, mostly of notational nature.  
\bigskip

\indent We first prove the estimate on the 1-point function, \eqref{dens}. Let $p_0=(x,m_i)$, with $x\in\mathbb R^3$ and $m_i\in\{1_a,1_b,2_a,2_b,3_a,3_b\}$. Recall the definition of the 1-point correlation function $\rho_1^{(q,\Lambda)}(p_0)$ in the state with $q$ boundary conditions, given in \eqref{eq:2.11}. Using \eqref{eq:generating}, we can write it as
 \begin{equation}
   \rho_1^{(q,\Lambda)} (p_0) =z\left.\frac{\delta}{\delta\tilde z(p_0)}\log Z(\Lambda|q)\right|_{\tilde z(p)\equiv z}=
    z\left.\frac{\delta}{\delta\tilde z(p_0)}\log Z^q(\Lambda)\right|_{\tilde z(p)\equiv z} +
    z\left.\frac{\delta}{\delta\tilde z(p_0)}\log \frac{Z(\Lambda|q)}{Z^{q} (\Lambda )}\right|_{\tilde z(p)\equiv z}
\end{equation}
 The Mayer expansion of the plate model implies that
\begin{equation}\label{eq:derivlnZ}
  z\left.\frac{\delta}{\delta\tilde z(p_0)}\log Z^q(A)\right|_{\tilde z(p)\equiv z}
  =\delta_{m,q}z(1+O(zk^2))
\end{equation}
for all finite $A\subset \mathbb R^3,$ uniformly in $A$, hence in particular for $A=\Lambda.$
The analogue of \eqref{dens} at finite volume follows from the following lemma. Eq.\eqref{dens} then follows from taking the limit $\Lambda \nearrow \mathbb R^3$,
which is easily obtained, using the uniform convergence of the Mayer and polymer expansions.

\begin{lemma}\label{le:mah}
Let $p_0$ be as above, $ q\in \{1,2,3 \}$ and $A\in  \mathfrak{Int}'$, see \eqref{int'}. If the constant $\bar \epsilon$  in \eqref{epsbar} is sufficiently small, then 
\begin{equation}\label{bound}
\left|  z\left.\frac{\delta}{\delta\tilde z(p_0)}\log \frac{Z(A|q)}{Z^{q} (A)}\right|_{\tilde z(p)\equiv z}
\right| \ \leq z\, O (\bar\epsilon^C)\mathds 1(A\ni x),
\end{equation}
for some $C>0$, uniformly in $A$. 
\end{lemma}

{\bf Proof.} We argue by induction on the size of $A$ or, more precisely, in the volume of $\bar A$, which is the smallest set in $\mathfrak{Int}$ containing $A$. 
If $\bar A$ is so small that $A$ cannot contain any contours, then 
$Z(A|q)=Z^{q} (A)$ and \eqref{bound} is trivially true.
Assume now by induction that  \eqref{bound} holds for all $a\in\mathfrak{Int}'$ such that $|\bar a|< |\bar A|$, and let us prove \eqref{bound}. 
By the analogue of Theorem~\ref{theorem:cluster} with $\Lambda$ replaced by $A\in{\mathfrak{Int}'}$ and plate-dependent activities,
\begin{eqnarray}
&&   z\left.\frac{\delta}{\delta\tilde z(p_0)}\log \frac{Z(A|q)}{Z^{q} (A)}\right|_{\tilde z(p)\equiv z} = \label{eq:6.4}\\
&&\qquad =\sum_{n\ge 0}\frac1{n!}\sum_{X_0,\ldots, X_n\in \mathfrak{B}^T(\bar A)}\phi^T(X_0,\ldots,X_n)
    z\left.\frac{\delta}{\delta\tilde z(p_0)}K_q^{(A)}(X_0)\right|_{\tilde z(p)\equiv z}\prod_{i=1}^nK_q^{(A)}(X_i).\nonumber
\end{eqnarray}
We claim that $\left.\frac{\delta}{\delta\tilde z(p_0)}K_q^{(A)}(X)\right|_{\tilde z(p)\equiv z}$ admits a bound similar to the one for $K_q^{(A)}(X)$, namely
\begin{equation} \big|\left.\frac{\delta}{\delta\tilde z(p_0)}K_q^{(A)}(X)\right|_{\tilde z(p)\equiv z}\big|\le \bar\epsilon^{\, c|X'|}e^{-m\,{\rm dist}'(X',\xi_{x})},\label{eq:6.5}\end{equation}
for some constants $c,m>0$, where $\xi_x$ is the center of the block containing $x$. Inserting \eqref{eq:6.5} in \eqref{eq:6.4}, together with $|K_q^{(A)}(X)|\le \bar\epsilon^{\,|X'|}$, the result follows. We are left with proving 
\eqref{eq:6.5}. 
\medskip

Recall that $K_q^{(A)}(X)=K_{q,1}^{(A)}(X)+K_{q,\ge 2}^{(A)}(X)$. We consider $K_{q,1}^{(A)}(X)$ first. Using the definition \eqref{K1}, we need to estimate 
 \begin{equation}
   \frac{\delta}{\delta\tilde z(p_0)} K_{q,1}^{(A)}(X):=
    \sum_{\mAthop{\gamma\in\mathcal C_1(\Lambda,q)}_{\Gamma_\gamma=X}}\frac{\delta}{\delta\tilde z(p_0)}\zeta_{q}^{(A)}(\gamma)
    \label{K1.1}
  \end{equation}
for $\tilde z(p_0)=z$. Recall that 
\begin{equation} \zeta_{q}^{(A)}(\gamma)=\Big(\frac{z^{|P_{\gamma}|}{\varphi (P_{\gamma })}}{Z^{q}(\Gamma_\gamma) }\Big)\Big(e^{-\int_{\Omega^q_A}dP\varphi^T(P)z^{|P|}F_\gamma(P)}\Big)\Big( \prod_{j=1}^{h_\Gamma}
    \frac{ Z^{(\gamma)} (\mathrm{Int}_{j}\Gamma_\gamma |m^j_{\mathrm{int},\gamma})}{Z(\mathrm{Int}_j\Gamma_\gamma |q )}\Big).\end{equation}
If the derivative hits the first parenthesis, then it can either act on $z^{|P_\gamma|}$, in which case it produces an indicator function $\mathds 1(P_\gamma\ni p_0)$, or on $Z^q(\Gamma_\gamma)$, in which case 
it generates an extra factor $\frac{\delta}{\delta\tilde z(p_0)}\log Z^q(\Gamma_\gamma)$, which by Mayer expansion is equal to $\mathds 1(x\in\Gamma_\gamma)\delta_{mq}(1+O(zk^2)).$
\medskip

\noindent If the derivative hits the second parenthesis, then it produces an extra factor $$-\frac{\delta}{\delta\tilde z(p_0)}\int_{\Omega^q_A} dP\ \varphi^T(P)z^{|P|}F_{\gamma}(P),$$ whose absolute value is bounded from above by 
$C(zk^2)^{\max\{0,\,c\,\mathrm{dist'}(\xi_x,\Gamma_\gamma')-1\}}\delta_{mq}$, for some $C,c>0$.

\noindent If the derivative hits the third parenthesis, we get extra factors of the form 
$$\frac{\delta}{\delta\tilde z(p_0)}\log\frac{Z(A_j^\gamma|m^j)}{Z^{m^j}(A_j^\gamma)}-
\frac{\delta}{\delta\tilde z(p_0)}\log\frac{Z(A_j|q)}{Z^{q}(A_j)}
+\frac{\delta}{\delta\tilde z(p_0)}\log\frac{Z^{m^j}(A_j^\gamma)}{Z^{q}(A_j)},$$
where, for short, we denoted $A_j={\rm Int}_j \Gamma_\gamma$, $m^j=m^j_{\mathrm{int},\gamma}$, $A_j^\gamma={\rm Int}_j \Gamma_\gamma\setminus V_{m^j}(P_\gamma)$ and $V_{m^j}(P_\gamma)$ is the excluded volume 
produced by the plates in $P_\gamma$. Now, the first two terms are estimated by the inductive assumption. The third term can be computed explicitly via Mayer expansion, and equals 
$\delta_{m,m^j}(1+O(zk^2))\mathds 1(x\in A_j^\gamma)-\delta_{m,q}(1+O(zk^2))\mathds 1(x\in A_j)$. 
\medskip

\noindent Putting things together we get 
\begin{equation} \big|\frac{\delta}{\delta\tilde z(p_0)}K_{q,1}^{(A)}(X)\big|\le C \bar\epsilon^{\, |X'|}e^{-m\,{\rm dist}'((X\cup_j{\rm Int}_jX)',\xi_{x})},\label{eq:6.5bis}\end{equation}
for some $C,m>0$. Now, note that, if $x\in \cup_j{\rm Int}_jX$, in which case ${\rm dist}'((X\cup_j{\rm Int}_jX)',\xi_{x})=0$, then $|X'|\ge c\, {\rm dist}'(X',\xi_{x})$ for some $c>0$, hence \eqref{eq:6.5} holds for $K_{q,1}^{(A)}(X)$. 
\bigskip

Finally, we consider $K_{q,\ge 2}^{(A)}(X)$. Using the definition \eqref{Kdef} we see that the derivative generates factors of the same form as above, plus an additional factor arising from the derivative of $\mathcal F_\partial(Y_j)$.
By repeating the strategy leading to \eqref{boundeF}, we get 
  \begin{equation}
    \left|\frac{\delta}{\delta\tilde z(p_0)}\mathcal F_\partial(Y_j)\right|
    \le  \delta_{mq} \mathds 1(x\in Y_j)  (C'zk^2)^{\max(2, c_0|Y'|)}\mathds 1_{\mathrm{dist}(Y,X_0)=0}.
   \end{equation}
This leads to the desired bound,  \eqref{eq:6.5}, for $K_{q,\ge 2}^{(A)}(X)$ and concludes the proof of the lemma. \qed
\bigskip

\indent The computation of the 2-point correlation function is quite similar: let $p_{1}= (x_{1},o_{1})$ $p_{2}= (x_{2},o_{2})$.
We write 
\begin{equation}
  \rho_{2}^{(q,\Lambda)}(p_{1},p_{2})-\rho_{1}^{(q,\Lambda )} (p_{1})\rho_{1}^{(q,\Lambda )} (p_{2}) =z^2\left.\frac{\delta^2}{\delta\tilde z(p_1)\delta\tilde z(p_2)}\log Z(\Lambda|q)\right|_{\tilde z(p)\equiv z}
\end{equation}
and $\log Z(\Lambda|q)=\log Z^q(\Lambda)+\log\frac{Z(\Lambda|q)}{Z^q(\Lambda)}$, so that the derivative produces 
two terms: the first is the second derivative of $\log Z^q(\Lambda)$, which we compute using the Mayer expansion, and the second is similar to the right side of \eqref{eq:6.4}, with 
two derivatives rather than one. These two derivatives have the effect of pinning the clusters of polymers to both $x_1$ and $x_2$ and, because of the exponential decay of their activity, this implies the exponential decay in~(\ref{rho2}). The details are left to the reader. This concludes the proof of Theorem \ref{theorem:nematic}.

\bigskip

\noindent{{\bf Acknowledgements.}} A.G. and M.D. acknowledge support from the European Research Council (ERC) under the European Union's Horizon 2020 research and innovation programme (ERC CoG UniCoSM, grant agreement n.724939).
The work of I.J. was supported by The Giorgio and Elena Petronio Fellowship Fund and The Giorgio and Elena Petronio Fellowship Fund II.

\end{document}